\documentclass[twocolumn,epjc3]{svjour3}   

\smartqed  

\usepackage[T1]{fontenc} 
\RequirePackage{graphicx,bm,amsfonts,hhline,lineno,hyperref,amsmath,amssymb,float,eurosym,latexsym,epsf,mathtools,cuted,makecell,array}
\usepackage[font={footnotesize,it}]{caption}
\usepackage[caption = false]{subfig}
\setcellgapes{4pt}

\journalname{Eur. Phys. J. C}

\begin{document}

\title{Anisotropic strange stars in the Einstein-Maxwell spacetime}

\author{Debabrata Deb\thanksref{e1,addr1}
\and Maxim Khlopov\thanksref{e2,addr2}
\and Farook Rahaman\thanksref{e3,addr3}
\and Saibal Ray\thanksref{e4,addr4}
\and B.K. Guha\thanksref{e5,addr1}.}

\thankstext{e1}{e-mail: ddeb.rs2016@physics.iiests.ac.in}
\thankstext{e2}{e-mail: khlopov@apc.in2p3.fr}
\thankstext{e3}{e-mail: rahaman@associates.iucaa.in}
\thankstext{e4}{e-mail: saibal@associates.iucaa.in}
\thankstext{e5}{e-mail: bkguhaphys@gmail.com}

\institute{Department of Physics, Indian Institute of Engineering Science
	   and Technology, Shibpur, Howrah, West Bengal, 711103, India\label{addr1} \and 
       National Research Nuclear University ``MEPHI'' (Moscow Engineering Physics Institute), 115409 Moscow, 
	   Russia Centre for Cosmoparticle Physics ``Cosmion'' 115409 Moscow, Russia and APC laboratory 10, rue Alice 
	   Domon et Lonie Duquet 75205 Paris Cedex 13, France \label{addr2} \and 
	   Department of Mathematics, Jadavpur University, Kolkata 700032, West Bengal, India \label{addr3} \and 
	   Department of Physics, Government College of Engineering and Ceramic Technology, Kolkata 700010, West Bengal, 
	   India \label{addr4} }

\date{Received: date / Accepted: date}

\maketitle

\begin{abstract}
We present here a detailed analysis on the effects of charge on the anisotropic strange star candidates by considering a spherically symmetric interior spacetime metric. To obtain exact solution of the Einstein-Maxwell field equations we have considered the anisotropic strange quark matter (SQM) distribution governed by the simplified MIT bag equation of state (EOS), $p=\frac{1}{3}\left( {\rho}-4\,B \right)$, where $B$ is the bag constant and the distribution of the electrical charge is given as $q(r)=Q\left({r}/{R}\right)^3=\alpha {r^3}$, where $\alpha$ is a constant. To this end, to calculate different constants we have described the exterior spacetime by the Reissner-Nordstr{\"o}m metric. By using the values of the observed mass for the different strange star candidates we have maximized anisotropic stress at the surface to predict the exact values of the radius for the different values of $\alpha$ and a specific value of the bag constant. Further, we perform different tests to study the physical validity and the stability of the proposed stellar model. We found accumulation of the electric charge distribution is maximum at the surface having electric charge of the order ${{10}^{20}}~C$ and electric field of the order ${10}^{21-22}~V/cm$. To study the different physical parameters and the effects of charge on the anisotropic stellar system we have presented our analysis graphically and in the tabular format by considering $LMC~X-4$ as the representative of the strange star candidates.    
\end{abstract}

\keywords{General Relativity; anisotropic fluid; strange stars.}

\maketitle

\section{Introduction}\label{sec1}
The theoretical possibility of the existence of hypothetical strange quark stars were first speculated in Refs.~\cite{Itoh1970,Farhi1984,Alcock1986,Haensel1986}. According to the strange quark matter hypothesis~\cite{Bodmer1971,Witten1984,Terazawa} the strange quark matter (SQM), made of equal number of up, down and strange quarks can be considered as the absolute ground state for the confined state of hadrons~\cite{Farhi1984,Alcock1986,Alcock1988,Madsen1999}.
Although strange stars form a distinct hypothetical branch of compact stars but these heavier members have masses and radii quite similar to the neutron stars. However, strange stars are not part of the the continuum of equilibrium configurations like white dwarf and neutron stars~\cite{Glendenning1995a,Glendenning1995b,Kettner1995}. In this context it is worth mentioning that strange matter equation of state (EOS) appears as the suitable EOS to explain observed compactness of the compact astrophysical objects like $4U~1820-30$, $SAX~J~1808.4-3658$, $4U~1728-34$, $Her~X-1$, $RX~J185635-3754$ and $PSR~0943+10$~\cite{Alcock1986,Haensel1986,Weber2005,Perez-Garcia2010,Rodrigues2011,Bordbar2011}, whereas neutron star EOS failed to explain those estimated compactness.

To maintain global charge neutrality strange stars which made of approximately equal number of up, down and strange quarks, should include smaller number of electrons. Alcock et al.~\cite{Alcock1986,Alcock1988} and Usov et al.~\cite{Usov2004,Usov2005} in their study showed that high electric fields in the order of ${10}^{18-19}~V/cm$ is expected to present on the surface of strange stars and presence of electrons play a significant role to the formation of the electric dipole layer at the surface. Such strong electric fields have values on the order of the energy density of SQM and it should be included in the stress-energy tensor which describes strange stars. The presence of the charge affects the relativistic stellar system in the following ways: (i) it causes the space-time curvature, (ii) it produces Coulomb interaction by introducing an extra term in the relativistic hydrodynamic equilibrium  equation, and (iii) the energy density associated with the electric field has significant role in producing the gravitational mass of the relativistic stellar system. In this line several literature~\cite{Ray2003,Siffert2007,Negreiros2009,Varela2010,Malheiro2011,Rahaman2012,Arbanil2013,Sunzu2014,Arbanil2015,Panahi2016} can be referred to understand the effects of the electric charge on the relativistic compact stellar system.

In his pioneering work Ruderman~\cite{Ruderman1972} first introduced the idea of pressure anisotropy and showed that the high density of the nuclear matters which interact relativistically are the key reason of the formation of the anisotropy. 
Here, by anisotropy we are addressing the difference between the radial component, ${p_r}(r)$ and the angular component, ${p_{\theta}}(r) = {p_{\phi}}(r) \equiv {p_{t}}(r)$ of the pressure. Clearly, ${p_{\theta}}(r) = {p_{\phi}}(r)$ is the consequence of the assumed spherical symmetry of the stellar system. An extensive study by Bowers and Liang~\cite{Bowers1974} showed that in the presence of complex strong interactions anisotropy in the spherically symmetric stellar system may be arising due to the presence of superconductivity and  superfluidity of the ultradense matter. Later, Herrera and Santos in their detailed review~\cite{Herrera1997} discussed the possible reasons behind the formation and existence of the local anisotropic stress in a self gravitating system and also studied their effect on a static spherically symmetric stellar system. Dev and Gleiser~\cite{Dev2002,Dev2003,Dev2004} in their series of work studied the significant effect of anisotropy on the redshift and maximum mass. They also showed that the presence of anisotropic stress enhances stability of the relativistic stellar system compared to the isotropic cases and  predicted that for the lower adiabatic index values too anisotropic systems are stable. In this line several authors as in Refs.~\cite{Ivanov2002,SM2003,MH2003,Usov2004,Varela2010,Rahaman2010,Rahaman2011,Rahaman2012,Kalam2012,Maurya2016,Maurya2017,Deb2017,Ovalle2017a,Ovalle2017b}  have studied the effect of anisotropy on the spherically symmetric compact stellar system and examined it's effect on the different physical properties of the stellar system.

The present work is the charged generalization of the earlier work done by Deb et al.~\cite{Deb2017}, where they presented an unique anisotropic model for the strange stars and showed the typical mass-radius relation for the strange stars by solving the Einstein field equations. To this end, they assumed simplified MIT bag EOS and showed that maximum anisotropy at the surface of the ultra dense strange stars is their inherent property. Using the motivation of the earlier work~\cite{Deb2017}, in the present article we have studied charged and anisotropic spherically symmetric stellar systems for the strange stars by considering a specific form of the electric charge distribution, $q(r)$. We also presented exact solutions for the Maxwell-Einstein field equations. It is interesting to note that though there are several literature which separately studied the effect of anisotropy or charge on the strange stars. But we found there is no other literature, which has studied the combined effects of anisotropy and charge on the compact stellar system by providing typical mass-radius relation for the strange stars in the framework of the Maxwell-Einstein gravity. However, in the present study by considering the combined effects of anisotropy and charge on the stellar system we attempt to present the exact solutions for the Maxwell-Einstein field equations by providing the typical mass-radius relation for the strange stars. Further, we have also examined the physical validity of the obtained solutions. 

The outline of our study is as follows: In Sect.~\ref{sec2} we have presented the basis of using the MIT bag EOS and the chosen form of the electric charge distribution. The basic equations to describe the anisotropic charged stellar system are presented in Sect.~\ref{sec3}. In Sect.~\ref{sec4} we have derived the solutions for the Maxwell-Einstein field equations and presented expressions for the different physical parameters. Further, in Sect.~\ref{sec5} to show physical acceptibility of the stellar system on the basis of the obtained solutions we have performed different tests like Energy conditions~\ref{subsec5.1}, mass-radius relation~\ref{subsec5.2}, compactification factor and redshift~\ref{subsec5.3}, and the stability of the system~\ref{subsec5.4}. Finally, in Sect.~\ref{sec6} we have concluded our study by discussing in detail the effects of the electric charge distribution on the anisotropic stellar system.

\section{The MIT Bag equation of state and the electric charge distribution}\label{sec2}
In the present article we consider MIT bag model EOS~\cite{Chodos1974} to describe the SQM distribution. in MIT bag model to maintain all the corrections due to energy and pressure functions of SQM an ad hoc bag function has been introduced. For the simplicity we assume that the up $(u)$, down $(d)$ and strange $(s)$ quarks are are massless and non-interacting in nature. Hence, the quark pressure, $p_r$ is defined as
 \begin{equation}\label{2.1}
{p_r}={\sum_{f=u,d,s}}{p^f}-{B},
 \end{equation}
where $p^f$ is the pressure due to individual quark flavors viz. $u$, $d$ and $s$. $B$ is the vacuum energy density and usually known as `Bag constant'. The relation between $p^f$ and energy density due to each quark flavors reads $p^f=\frac{1}{3}{{\rho}^f}$. Hence the energy density, $\rho$ due to de-confined SQM distribution inside the bag is defined as
\begin{equation}
{{\rho}}={\sum_{f=u,d,s}}{{\rho}^f}+B, \label{2.2}
\end{equation}
Hence, substituting relation between $p^f$ and ${{\rho}^f}$ into Eq.~(\ref{2.2}) and using Eq.~(\ref{2.1}) we have the simplified form of the MIT bag model EOS given as 
\begin{equation}
{p_r}=\frac{1}{3}({{\rho}}-4\,B).\label{2.3}
\end{equation} 
 In the recent times applying this simplified form of the MIT bag EOS several authors successfully studied strange star model~\cite{Negreiros2009,Malheiro2011,Rahaman2012,Brilenkov2013,Maharaj2014,Paulucci2014,Sunzu2014,Abbas2015,Isayev2015,Panda2015,Arbanil2015,Arbanil2016,Panahi2016,Deb2017,Lugones2017}.
 Following Rahaman et al.~\cite{Rahaman2014} we consider the value of the bag constant as $B=83~MeV/{{fm}^3}$.
 
To study the effects of charge on the relativistic stellar system  Felice et al.~\cite{Felice1995,Felice1999} in their literature considered an specific form of electric charge distribution $q(r)$ given as $q(r)=Q(r/R)^n$. Following Felice et al.~\cite{Felice1995,Felice1999} in the present study we choose this specific simplest form of $q(r)$, for the parametric values of $n=3$ as follows
\begin{eqnarray}\label{2.4}
q\left(r\right)=Q\left( \frac{r}{R} \right)^3\equiv \alpha\,{r^3},
\end{eqnarray}
where $Q$ and $R$ are the total charge and the total radius of the stellar system, respectively and $\alpha$ is a constant which can be defied as $\alpha=Q/R^3$.

\section{Basic stellar structure equations}\label{sec3}
To describe interior spacetime of the ultra dense spherically symmetrical stellar system in Schwarzschild-like coordinates~\cite{Tolman1939,Oppenheimer1939} we use metric as follows
\begin{equation}\label{3.1}
ds^2=e^{\nu(r)}dt^2-e^{\lambda(r)}dr^2-r^2(d\theta^2+\sin^2\theta d\phi^2),
\end{equation}
where the metric potential $\nu$ and $\lambda$ are the functions of the radial coordinate $r$ only. Now, to obtain hydrostatic stellar structure of the charged sphere we have to solve the Einstein-Maxwell field equations provided as
\begin{equation}
 R^i_j - \frac{1}{2} R\, {g^i_j} = 8\pi \left({T^i_j} + {E^i_j}\right), \label{3.2}
\end{equation}
where we assumed $G=1=c$ in the relativistic geometrized unit. Here $T^i_j$ and $E^i_j$ represents stress-energy tensor for the locally anisotropic fluid distribution and the present electromagnetic field, respectively and they are defined as~\cite{Dionysiou1982}
\begin{eqnarray}\label{3.3}
&\qquad {T^i_j} = \left[(\rho + p_t)v^iv_j - p_t{\delta^i}_j + (p_r - p_t) u^i u_j\right],\\ \label{3.4}
&\qquad\hspace{-0.8cm} {E^i_j} = \frac{1}{4\pi}\left(-F^{im}F_{jm} + \frac{1}{4\pi}{\delta^i}_jF^{mn}F_{mn}\right),
\end{eqnarray}
where $v^i$ and $u^i$ are the four-velocity and radial four-vector, respectively, $\rho$ is the energy density of SQM, $p_r$ represents pressure in the direction of $u^i$, known as radial pressure and $p_t$ represent component of pressure normal to  $u_i$, known as tangential pressure. Here, $F_{ij}$ is the anti-symmetric electromagnetic field tensor and can be defined as
\begin{equation}\label{3.5}
F_{ij} = \frac{\partial A_j}{\partial x_i} - \frac{\partial A_i}{\partial x_j},
\end{equation}
where, $A_j=(\phi(r), 0, 0, 0)$  is the four-potential. $F_{ij}$ satisfies the covariant Maxwell equations,
\begin{eqnarray} \label{3.6}
&\qquad F_{ik,j} + F_{kj,i} + F_{ji,k} = 0,\\ \label{3.7}
&\qquad \left[{\sqrt -g} F^{ik}\right]_{,k} = -4\pi J^i {\sqrt -g},
\end{eqnarray}
where  $J^i$ is the electromagnetic four-current vector defined as
\begin{equation}
J^i = \frac{\sigma}{\sqrt g_{44}} \frac{dx^i}{dx^4} = \sigma v^i,\label{3.8}
\end{equation}
where $\sigma=e^{\nu /2} J^{0}\left(r\right)$ represents the charged density and $g$ is the determinant of the metric $g_{ij}$ defined by 
\begin{equation}\label{3.9}
g = \left(\begin{array}{cccc} e^{\nu} & 0           & 0    & 0\\
                              0       &-e^{\lambda} & 0    & 0\\
                              0       & 0           & -r^2 & 0\\
                              0       & 0           & 0    & -r^2sin^2\theta \end{array} \right) = - e^{\nu+\lambda}r^4 sin^2\theta.
\end{equation}

For a static spherically symmetric stellar system $J^0$ is the only non vanishing component of the the electromagnetic four-current $J^i$ which is a function radial coordinate, r only. $F^{01}$ and $F^{10}$ are the only non zero components of the electromagnetic field tensor and they are related by $F^{01} = - F^{10}$.  $F^{01}$ and $F^{10}$ are the radial component of the electric field. Using Eqs.~(\ref{3.7}) and (\ref{3.8}) the expression for the electric field is given as
\begin{eqnarray}\label{3.10}
E\left(r\right)=F^{01}\left(r\right)=\frac{1}{r^2} e^{-\left(\nu+\lambda\right)/2} 4\pi\,\int_0^r {r^{\prime 2}} \sigma e^{\lambda /2} dr^{\prime}.
\end{eqnarray} 

If $q(r)$ represents the total charge of a spherical system of radius $r$ then following the relativistic Gauss's law the electric charge $q(r)$ can be defined as 
\begin{equation} \label{3.11}
q(r) = 4\pi \int_0^r \sigma r^{\prime 2} e^{\lambda/2} dr^{\prime} = r^2 \sqrt{-F_{14}F^{14}}. 
\end{equation}

Using Eqs.~(\ref{3.3}), (\ref{3.4}), (\ref{3.7}), (\ref{3.8}), (\ref{3.10}) and (\ref{3.11}) the stress-energy tensor for the anisotropic charged matter distribution can be written as
\begin{equation}\label{3.12}
T^{a}_{b} = \left(\begin{array}{cccc} -\left( \rho+\frac{q^2}{8\pi r^4} \right) & 0           & 0    & 0\\
                              0       &  p_r-\frac{q^2}{8\pi r^4}   & 0    & 0\\
                              0       & 0           &  p_t+\frac{q^2}{8\pi r^4}  & 0\\
                              0       & 0           & 0    &  p_t+\frac{q^2}{8\pi r^4}  \end{array} \right),
\end{equation}
where the electric charge and the electric field are related by $q^2(r)/8\pi r^4=E^2(r)/8\pi$.

Substituting Eq.~(\ref{3.12}) into Eq.~(\ref{3.2}) we have the explicit form of the Einstein field equation for the anisotropic charged spherically symmetric stellar system as follows~\cite{Dionysiou1982}
\begin{eqnarray}\label{3.13}
&\qquad\hspace{-1cm} {{\rm e}^{-\lambda }} \left( {\frac {\lambda^{{\prime}}}{r}}-\frac{1}{r^2}\right) +\frac{1}{r^2}=8\pi   \rho+{E}^{2},\\ \label{3.14}
&\qquad\hspace{-1cm} {{\rm e}^{-\lambda}} \left( {\frac {\nu^{{\prime}}}{r}}+\frac{1}{r^2}\right) -\frac{1}{r^2}=8\pi  p_{{r}}-{E}^{2},\\ \label{3.15}
&\qquad\hspace{-0.6cm} \frac{{\rm e}^{-\lambda}}{2} \left( \nu^{{\prime\prime}}+\frac{{\nu^{{\prime}}}^{2}}{2}+{\frac {\nu^{{\prime}}-\lambda^{{\prime}}}{r}}-\frac{\nu^{{\prime}}\lambda^{{\prime}}}{2} \right) =8\pi p_{{t}}+{E}^{2}.
\end{eqnarray}

In the analogy of the uncharged case, let we define the mass function of the spherically symmetric charged stellar system as follows
\begin{eqnarray}\label{3.16}
&\qquad m \left( r \right) =4\pi \int_{0}^{r}{{\rho}_{eff}} \left( r \right) {r}^{2}{dr}=4\pi \int_{0}^{r}\left({{\rho}}+\frac{E^2}{8\pi}\right) {r}^{2}{dr},\nonumber \\
\end{eqnarray}
where ${{\rho}_{eff}}={{\rho}}+\frac{E^2}{8\pi}$.

To describe the exterior spacetime of our system we consider the exterior Reissner-Nordstr{\"o}m metric given as
\begin{eqnarray}\label{3.17}
&\qquad\hspace{-1cm} ds^2 = \left(1 - \frac{2M}{r} +\frac{Q^2}{r^2}\right) dt^2- \frac{1}{\left(1 - \frac{2M}{r} + \frac{Q^2}{r^2}\right)} dr^2\nonumber\\
&\qquad\hspace{3.2cm} - r^2(d\theta^2 + sin^2\theta d\phi^2).
\end{eqnarray}

Now using Eqs. (\ref{2.4}), (\ref{3.16}) and (\ref{3.17}) we find from Eqs.~(\ref{3.13}) as follows
\begin{eqnarray}\label{3.18}
{{\rm e}^{-\lambda \left( r \right) }}=1-{\frac {2m}{r}}+{\frac {{q}^{2}}{{r}^{2}}}.
\end{eqnarray}

Following Mak and Harko~\cite{Harko2002} to obtain singularity free monotonically decreasing SQM density function $\rho$, we define 
\begin{equation}\label{3.19}
\rho(r)=\rho_c\left[1-\left(1-\frac{\rho_0}{\rho_c}\right)\frac{r^{2}}{R^{2}}\right],
\end{equation}
where ${\rho}_c$ and ${\rho}_0$ are the central and surface density, respectively.

To obtain hydrostatic equilibrium equation for the anisotropic charged stellar system we perform covariant divergence of the electromagnetic stress-energy tensor, i.e., ${\nabla}^a {T^a_b}=0$, which leads to the equation of energy conservation as follows
\begin{eqnarray}\label{3.20}
\hspace{0cm} {\frac {{\rm d}p_{{r}}}{{\rm d}r}} =-\left( {\rho+p_{{r
}}}\right)\frac{\left[ m+4\,\pi \,{r}^{3} \left( p_{{r}}-{\frac {{q}
^{2}}{4\pi {r}^{4}}} \right)  \right]} {{{r}^{2}}\left( 1-{\frac {2m}{r}}+{\frac {{q}^{2}}{{r}^{2}}} \right)}+\frac {q}{4\pi {r}^{4}}{{\frac {{\rm d}q}{{\rm d}r}}}\nonumber \\
&\qquad\hspace{-4cm} +\frac {2}{r}\left(p_{{t}}-p_{{r}}\right). 
\end{eqnarray}

For $q=0$ in Eq.~(\ref{3.20}) we retrieve the usual form of the Tolman-Oppenheimer-Volkoff (TOV) equation for the anisotropic matter distribution.

\section{Solution of the Maxwell-Einstein field equations}\label{sec4}
Using Eqs.(\ref{2.3}),~(\ref{2.4}),~(\ref{3.16}),~(\ref{3.18}) and (\ref{3.19}) and by solving the Maxwell-Einstein field equations (\ref{3.13})-(\ref{3.14}) we derive expression for the different physical parameters as follows
\begin{eqnarray}\label{4.1}
&\qquad\hspace{-1cm} \lambda=-\ln  \left\lbrace 1+ {\frac {\left(\lambda_{{2}}+\lambda_{{1}
}{r}^{2}\right)r^2}{{R}^{5}}} \right\rbrace,  \\ \label{4.2}
&\qquad\hspace{-1cm} \nu={\frac {1024}{3\,\nu_{{1}}\nu_{{4}}}} \Bigg[ \nu_{{2}}{\rm arctanh} \left({\frac {-4\,{R}^{5}{\alpha}^{2}{r}^{2}-32\,B\pi \,{R}^{3}{r}^{2}+6\,M{r}^{2}-\nu_{{5}}}{{R}^{2}\nu_{{1}}}}
\right)\nonumber\\
&\qquad\hspace{-1.5cm}+\nu_{{2}}{\rm arctanh} \left({\frac {{R}^{5}{\alpha}^{2}+16 B\pi {R}^{3}-M}{\nu_{{1}}}}\right)\nonumber\\
&\qquad\hspace{1cm}-{\frac {3 \nu_{{1}} \left\lbrace \nu_{{3}}\ln  \left( -{R}^{5}+\nu_{{4}}{r}^{4}+\nu_{{5}}{r}^{2} \right) - \nu_{{6}} \right\rbrace}{64}}\Bigg],  \\ \label{4.3}
&\qquad\hspace{-1cm} \rho=-{\frac {9\,{R}^{7}{\alpha}^{2}-9\,{R}^{5}{\alpha}^{2}{r}^{2
}+48\,B\pi \,{R}^{5}-80\,B\pi \,{R}^{3}{r}^{2}-15\,M{R}^{2}+15\,M{r}^{2}}{8\pi {R}^{5}}}, \nonumber \\ \\ \label{4.4}
&\qquad\hspace{-1cm} p_{{r}}=-{\frac {9\,{R}^{7}{\alpha}^{2}-9\,{R}^{5}{\alpha}^{2}{r
}^{2}+80\,B\pi \,{R}^{5}-80\,B\pi \,{R}^{3}{r}^{2}-15\,M{R}^{2}+15\,M{r}^{2}}{24\pi {R}^{5}}},\nonumber \\ \\ \label{4.5}
&\qquad\hspace{-1cm} p_t= \Big[\Big( 81\,{R}^{14}{\alpha}^{4}{r}^{2}-135\,{R}^{12}{\alpha}^{4}{r}^{4}+54\,{
R}^{10}{\alpha}^{4}{r}^{6}\nonumber\\
&\qquad\hspace{-1cm}+1008\,B\pi \,{R}^{12}{\alpha}^{2}{r}^{2}-1632\,B\pi \,{R}^{10}{\alpha}^{2}{r}^{4}+720\,B\pi \,{R}^{8}{\alpha}^{2}{r}^{6}\nonumber\\
&\qquad\hspace{-1cm}+3328\,{B}^{2}{\pi }^{2}{R}^{10}{r}^{2}-5888\,{B}^{2}{\pi }^{2}{R}^{8}{r}^{4}+2560\,{B}^{2}{\pi }^{2}{R}^{6}{r}^{6}\nonumber\\
&\qquad\hspace{-1cm}-270\,M{R}^{9}{\alpha}^{2}{r}^{2}+360\,M{R}^{7}{\alpha}^{2}{r}^{4}-135\,M{R}^{5}{\alpha}^{2}{r}^{6}\nonumber\\
&\qquad\hspace{-1cm}-27\,{R}^{12}{\alpha}^{2}+27\,{R}^{10}{\alpha}^{2}{r}^{2}-1680\,BM\pi \,{R}^{7}{r}^{2}\nonumber\\
&\qquad\hspace{-1cm}+2304\,BM\pi \,{R}^{5}{r}^{4}-960\,B
M\pi \,{R}^{3}{r}^{6}-240\,B\pi \,{R}^{10}\nonumber\\
&\qquad\hspace{-1cm}+480\,B\pi \,{R}^{8}{r}^{2}+225\,{M}^{2}{R}^{4}{r}^{2}-225\,{M}^{2}{R}^{2}{r}^{4}+90\,{M}^{2}{r}^{6}\nonumber\\
&\qquad\hspace{-3cm}+45\,M{R}^{7}-90\,M{R}^{5}{r}^{2} \Big)\Big/72\,{R}^{5}\pi \, \Big( 3\,{R}^{7}{\alpha}^{2}{r}^{2}\nonumber\\
&\qquad\hspace{-1cm}-2\,{R}^{5}{\alpha}^{2}{r}^{4}+16\,B\pi \,{R}^{5}{r}^{2}-16\,B\pi \,{R}^{3}{r}^{4}
-5\,M{R}^{2}{r}^{2}\nonumber\\
&\qquad\hspace{-1cm}+3\,M{r}^{4}+{R}^{5} \Big) \Big],
\end{eqnarray}
where $\lambda_{{1}}$,~$\lambda_{{2}}$,~$\nu_{{1}}$,~$\nu_{{2}}$,~$\nu_{{3}}$,~$\nu_{{4}}$,~$\nu_{{5}}$~and~$\nu_{{6}}$ are constants and their expressions are shown in Appendix~\ref{App}. 

\begin{figure}[!htp]\centering
	\includegraphics[scale=.3]{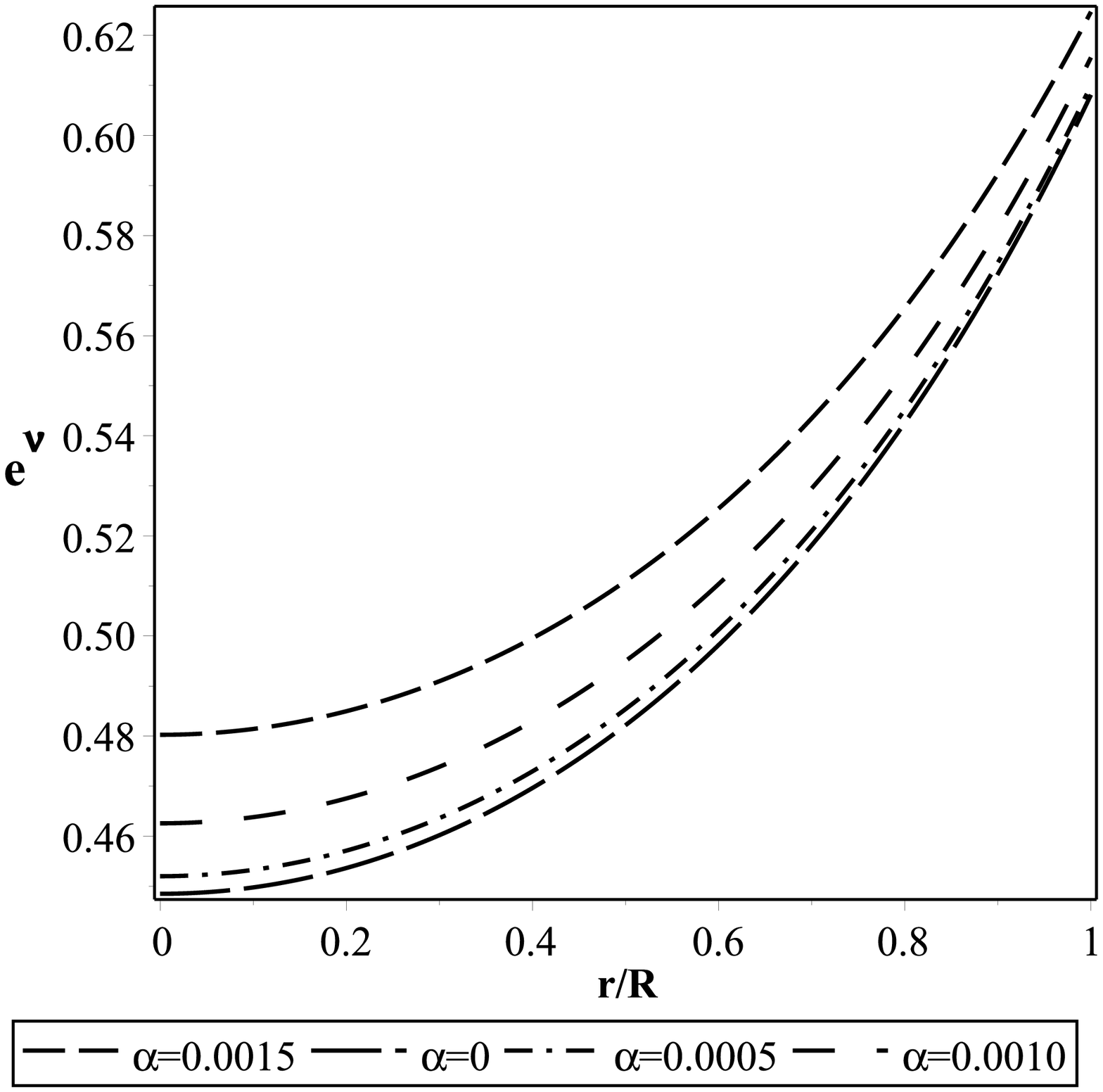}
	\includegraphics[scale=.3]{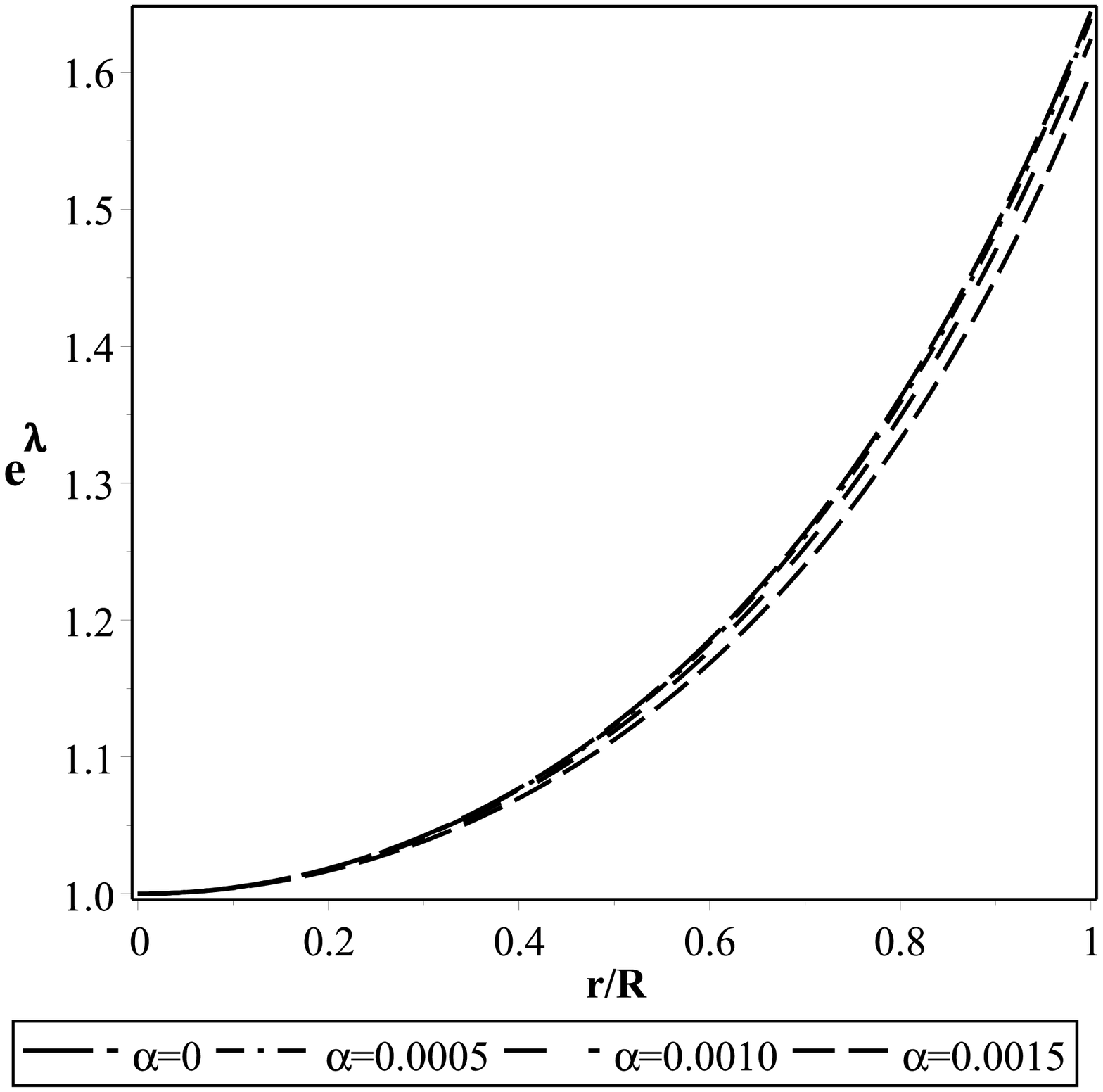}
		\caption{Variation of (i) ${e}^{\nu(r)}$ (upper panel) and (ii) ${e}^{\lambda(r)}$ (lower panel) as a function of the radial coordinate $r/R$ for the strange star $LMC~X-4$. Here $B=83~ MeV/{{fm}^3}$} \label{Fig1}
\end{figure}

\begin{figure}[!htp]\centering
          \includegraphics[scale=.3]{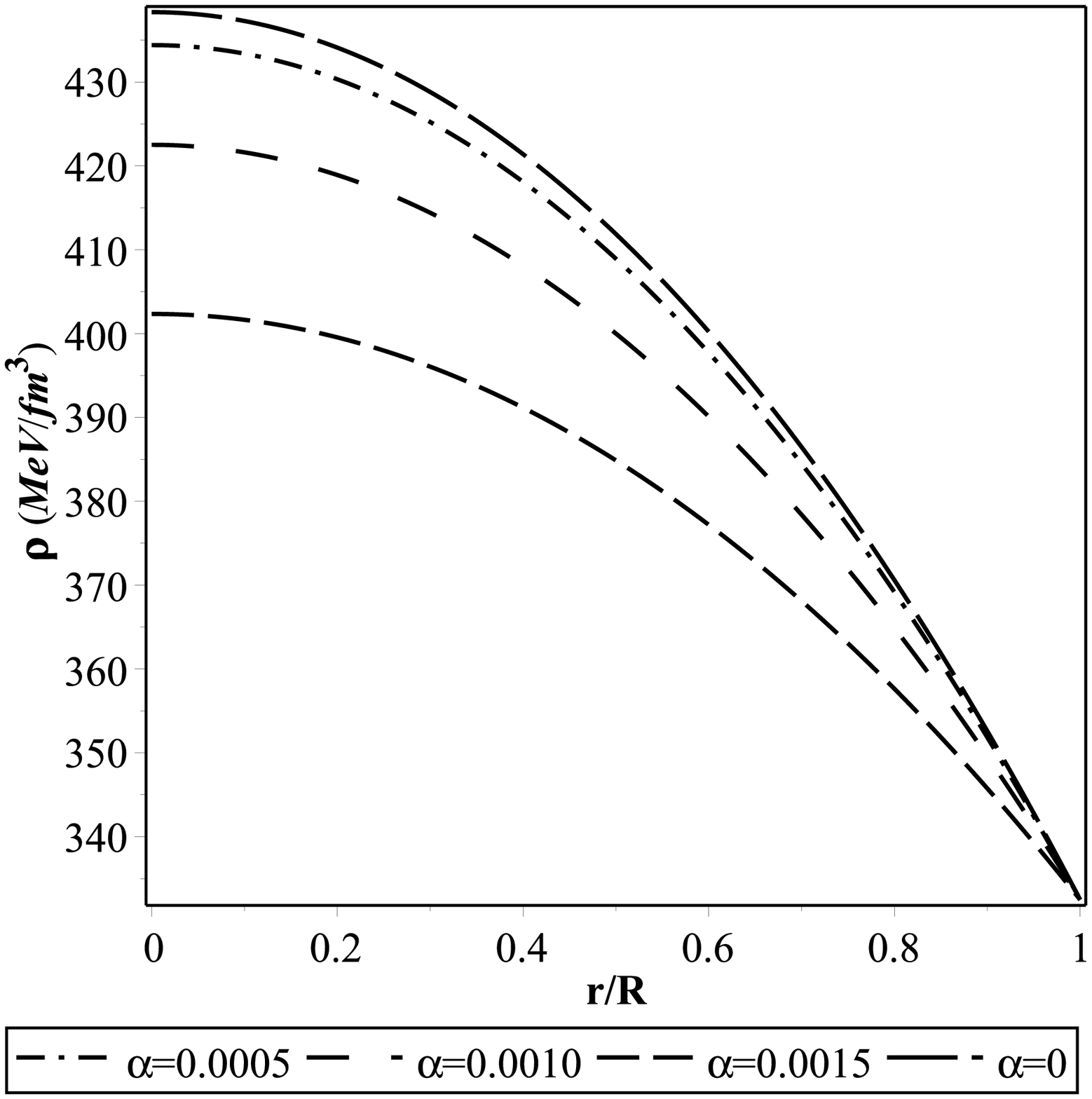}
          \includegraphics[scale=.3]{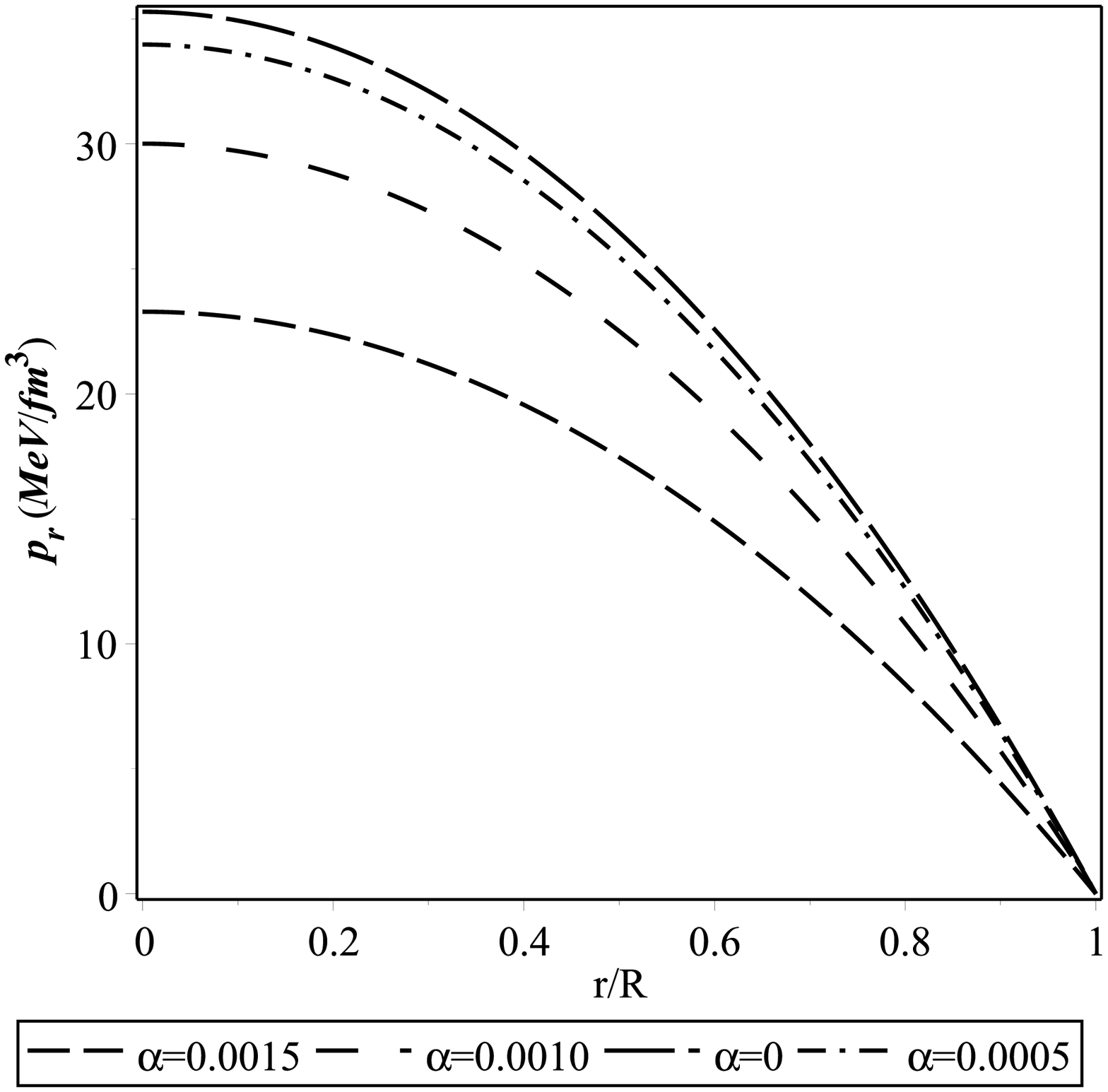}
          \includegraphics[scale=.3]{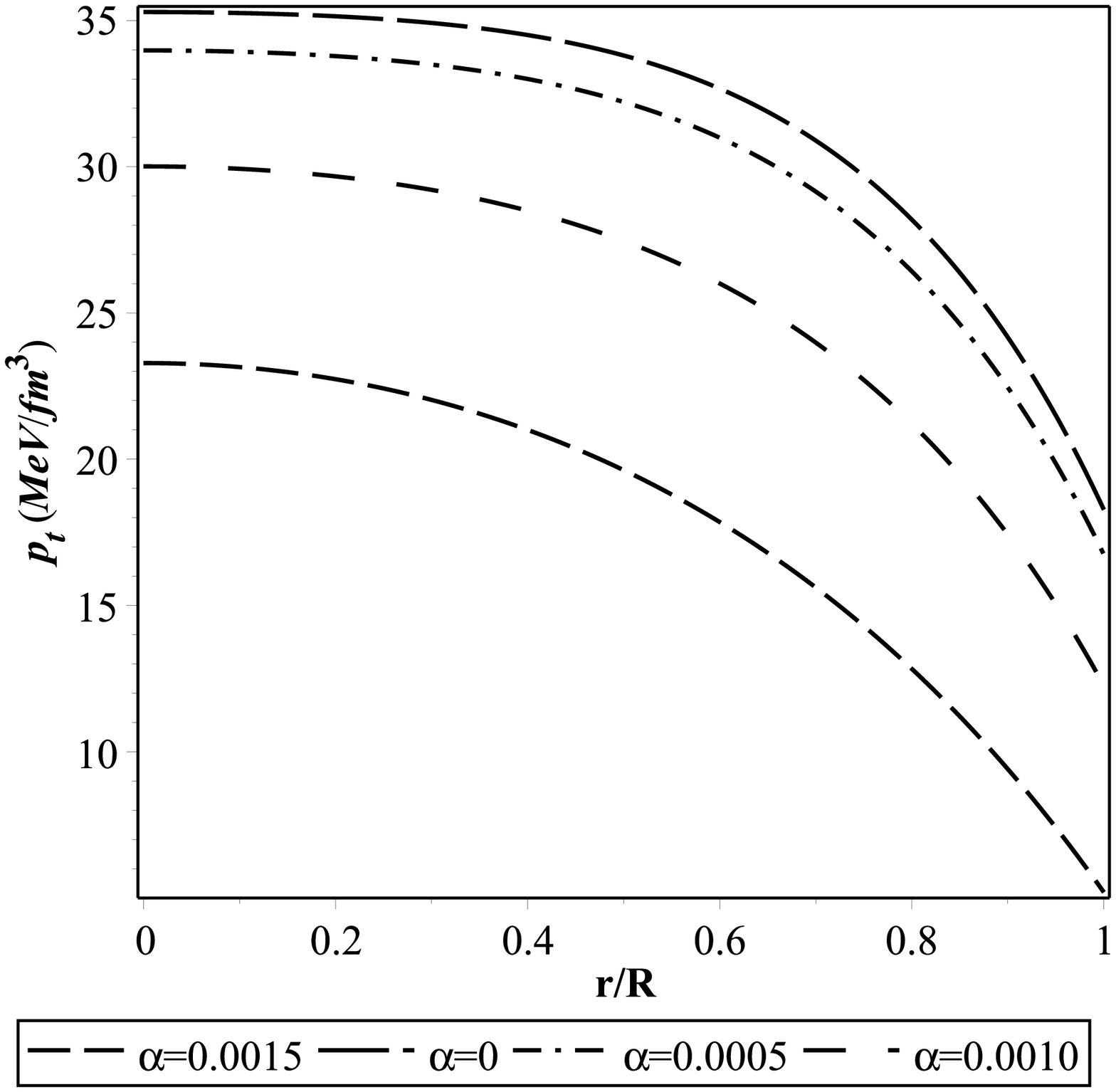}
\caption{Variation of (i) $\rho$ (upper panel), (ii) $p_{{{r}}}$ (middle panel) and (iii) $p_{{{t}}}$ (lower panel) as a function of the radial coordinate $r/R$ for the strange star $LMC~X-4$} \label{Fig2}
\end{figure}

We featured variation of the physical parameters, viz. $e^{\lambda}$,~$e^{\nu}$,~$\rho$,~$p_r$ and $p_t$ with respect to the radial coordinate $r/R$ in Figs.~\ref{Fig1} and \ref{Fig2}.


The anisotropy for our system is given as
\begin{eqnarray}\label{4.6}
&\qquad\hspace{-1cm} \Delta=p_t-p_r=\Bigg[{r}^{2} \Big( 162\,{R}^{14}{\alpha}^{4}-270\,{R}^{12}{\alpha}^
{4}{r}^{2}\nonumber\\
&\qquad\hspace{-1cm}+108\,{R}^{10}{\alpha}^{4}{r}^{4}+2160\,B\pi \,{R}^{12}{\alpha}^{2}-3696\,B\pi \,{R}^{10}{\alpha}^{2}{r}^{2}\nonumber\\
&\qquad\hspace{-1cm}+1632\,B\pi \,{R}^{8}{\alpha}^{2}{r}^{4}+7168\,{B}^{2}{\pi }^{2}{R}^{10}-13568\,{B}^{2}{\pi }^{2}{R}^{8}{r}^{2}\nonumber\\
&\qquad\hspace{-1cm}+6400\,{B}^{2}{\pi }^{2}{R}^{6}{r}^{4}-540\,M{R}^{9}{\alpha}^{2}+801\,M{R}^{7}{\alpha}^{2}{r}^{2}\nonumber\\
&\qquad\hspace{-1cm}-306\,M{R}^{5}{\alpha}^{2}{r}^{4}-3600\,BM\pi \,{R}^{7}+5664\,BM\pi \,{R}^{5}{r}^{2}\nonumber\\
&\qquad\hspace{-1cm}-2400\,BM\pi \,{R}^{3}{r}^{4}+240\,B\pi \,{R}^{8}+450\,{M}^{2}{R}^{4}\nonumber\\
&\qquad\hspace{-1cm}-585\,{M}^{2}{R}^{2}{r}^{2}+225\,{M}^{2}{r}^{4}-45\,M{R}^{5} \Big) \Big/ 72\,{R}^{5}\pi \Big( 3\,{R}^{7}{\alpha}^{2}{r}^{2}\nonumber\\
&\qquad\hspace{-1cm}-2\,{R}^{5}{\alpha}^{2}{r}^{4}+16\,B\pi \,{R}^{5}{r}^{2}-16\,B\pi \,{R}^{3}{r}^{4}
-5\,M{R}^{2}{r}^{2}\nonumber\\
&\qquad\hspace{-1cm}+3\,M{r}^{4}+{R}^{5} \Big) \Bigg].
\end{eqnarray}

\begin{figure}[!htp]\centering
	\includegraphics[scale=.3]{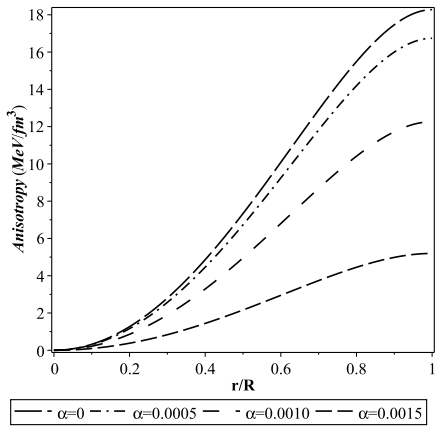}
			\caption{Variation of anisotropy as a function of the radial coordinate $r/R$ for the strange star $LMC~X-4$} \label{Fig3}
\end{figure}

The variation of the anisotropic stress $\left(\Delta\right)$ with respect to the radial coordinate $r/R$ is shown in Fig.~\ref{Fig3}. We find from Fig.~\ref{Fig3} anisotropy is zero at the center and maximum at the surface as predicted by Deb et al.~\cite{Deb2017}.


Now, using the observed values of the mass of the different strange star candidates as shown in Table 1 and following Deb et al.~\cite{Deb2017} we shall maximize the anisotropic stress, $\Delta(r)$ at the surface $r=R$ to predict the exact value of the radius, $R$ for the different strange stars. To this end, we consider that the value of the bag constant is $B=83~ MeV/{{fm}^3}$~\cite{Rahaman2014} and the chosen values of $\alpha$ are $0$, $0.0005$, $0.0010$ and $0.0015~{{km}^{-2}}$. Clearly, ${\Delta}^{\prime}(R)=0$ will yield several values of $R$ and we will choose only that value of $R$ for which the Buchdahl conditions~\cite{Buchdahl1959} will be satisfied.

\begin{figure}[!htp]\centering
	\includegraphics[scale=.3]{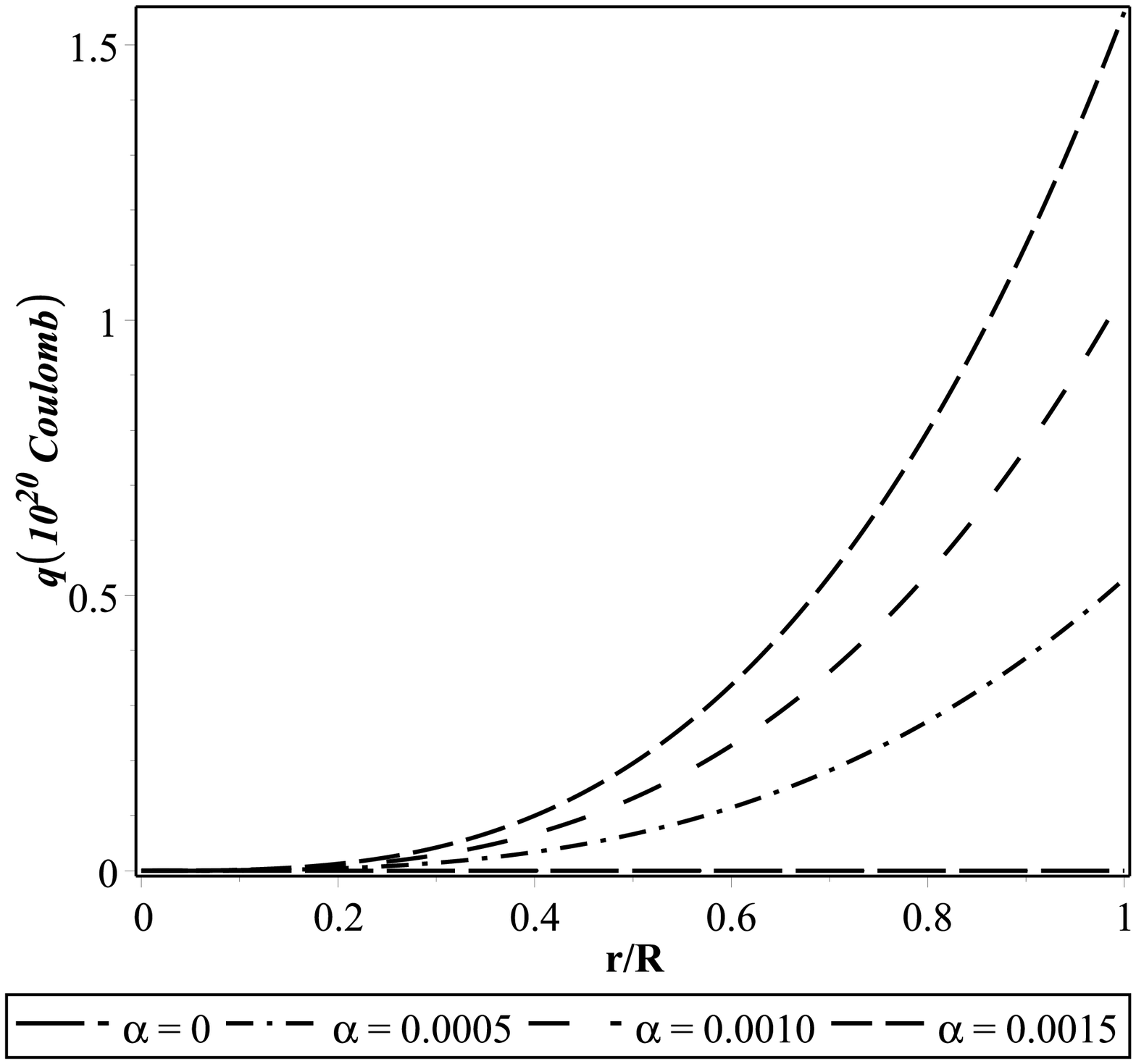}
	\includegraphics[scale=.3]{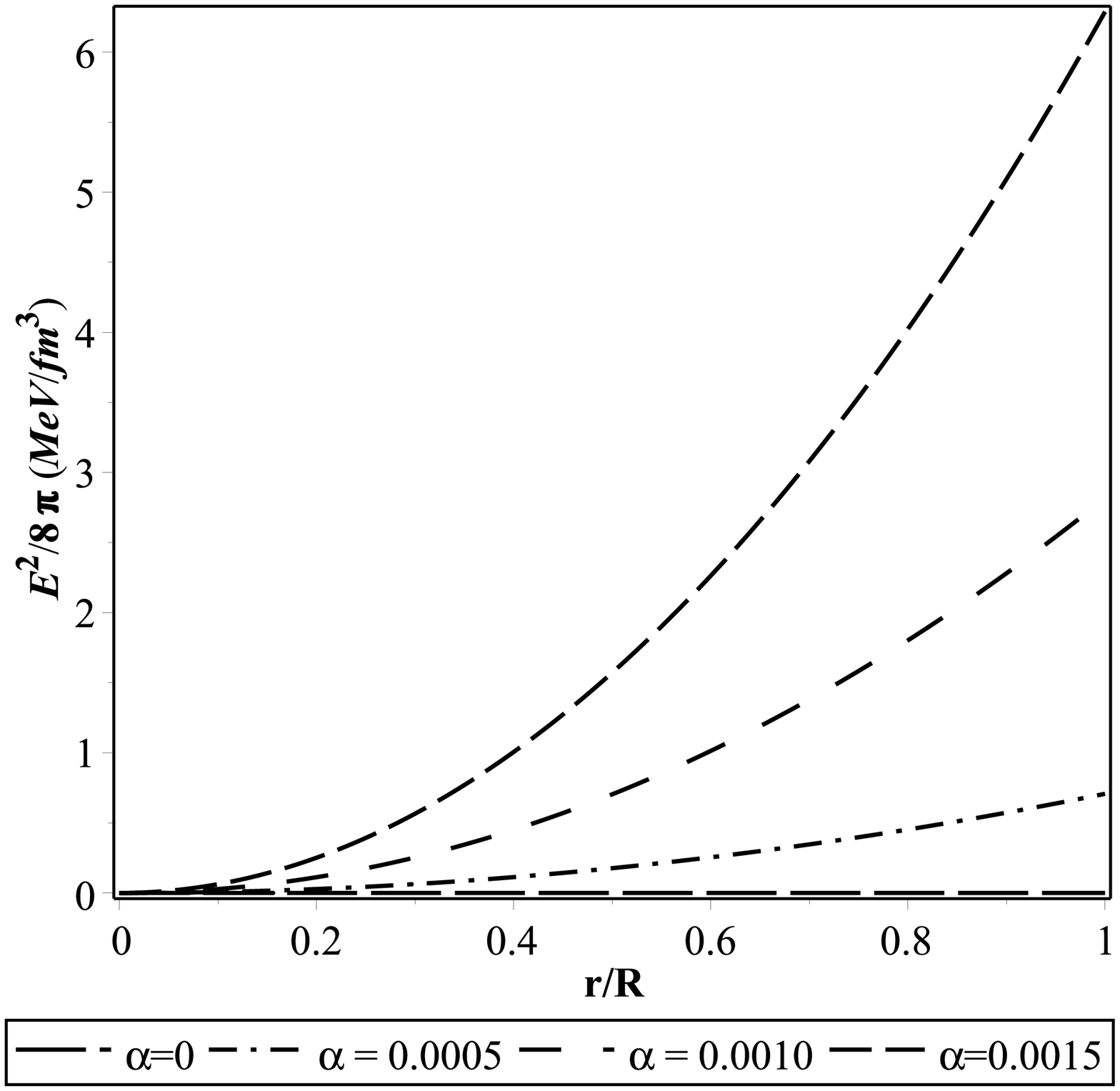}
			\caption{Variation of (i) $q\left(r\right)$ (upper panel) and (ii) $E^2 \left(r\right)/8\pi$ (lower panel) as a function of the radial coordinate $r/R$ for the strange star $LMC~X-4$} \label{Fig4}
\end{figure}

In Fig.~\ref{Fig4} we have featured the variation of the electric charge distribution $q(r)$ and electrical energy density $E^2(r)/8\pi$ with respect to the radial coordinate $r/R$ in the upper and lower panel, respectively. Fig.~\ref{Fig4} clearly suggests that both the distribution of the electric charge and electrical energy density is minimum, i.e., zero at the center and maximum at the surface.


\section{Salient physical features of the anisotropic charged stellar system}\label{sec5}
In this section to discuss physical validity of the achieved solution we will study some salient physical features of the stellar system as follows:

\subsection{Energy conditions}\label{subsec5.1}
To satisfy energy conditions, viz., Null Energy Condition (NEC),Weak Energy Condition (WEC), Strong Energy Condition (SEC) and Dominant Energy Condition (DEC) the anisotropic charged stellar system have to be consistent with all the inequalities simultaneously as follows
\begin{eqnarray}\label{5.1.1}
&\qquad\hspace{-0.5cm}~NEC:\rho+p_r\geq 0,~\rho+p_t+\frac{E^2}{4\pi}\geq 0, \\ \label{5.1.2}
&\qquad\hspace{-1cm}~WEC: \rho+p_r\geq 0,~\rho+\frac{E^2}{8\pi}\geq 0,~\rho+p_t+\frac{E^2}{4\pi}\geq 0, \\ \label{5.1.3}
&\qquad\hspace{-0.5cm}~SEC: \rho+p_r\geq 0,~\rho+p_r+2\,p_t+\frac{E^2}{4\pi}\geq 0, \\ \label{5.1.4}
&\qquad\hspace{-0.5cm}~DEC: {\rho}+\frac{E^2}{8\pi}\geq 0, {{\rho}-{p_r}}+\frac{E^2}{4\pi}\geq 0, {{\rho}-{p_t}}\geq 0. 
\end{eqnarray}

\begin{figure}[!htpb]
\centering
   \includegraphics[scale=0.2]{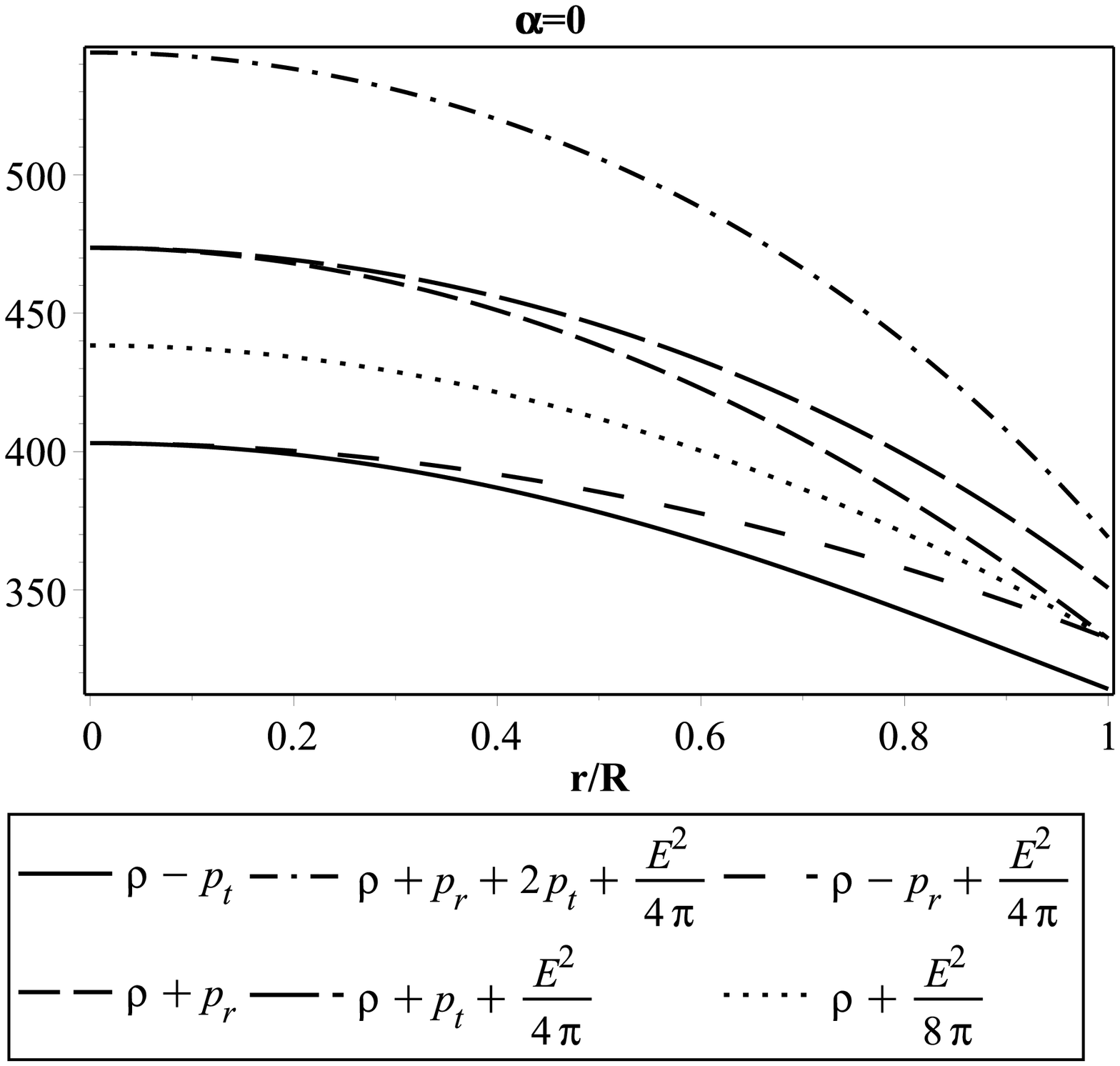}
   \includegraphics[scale=0.2]{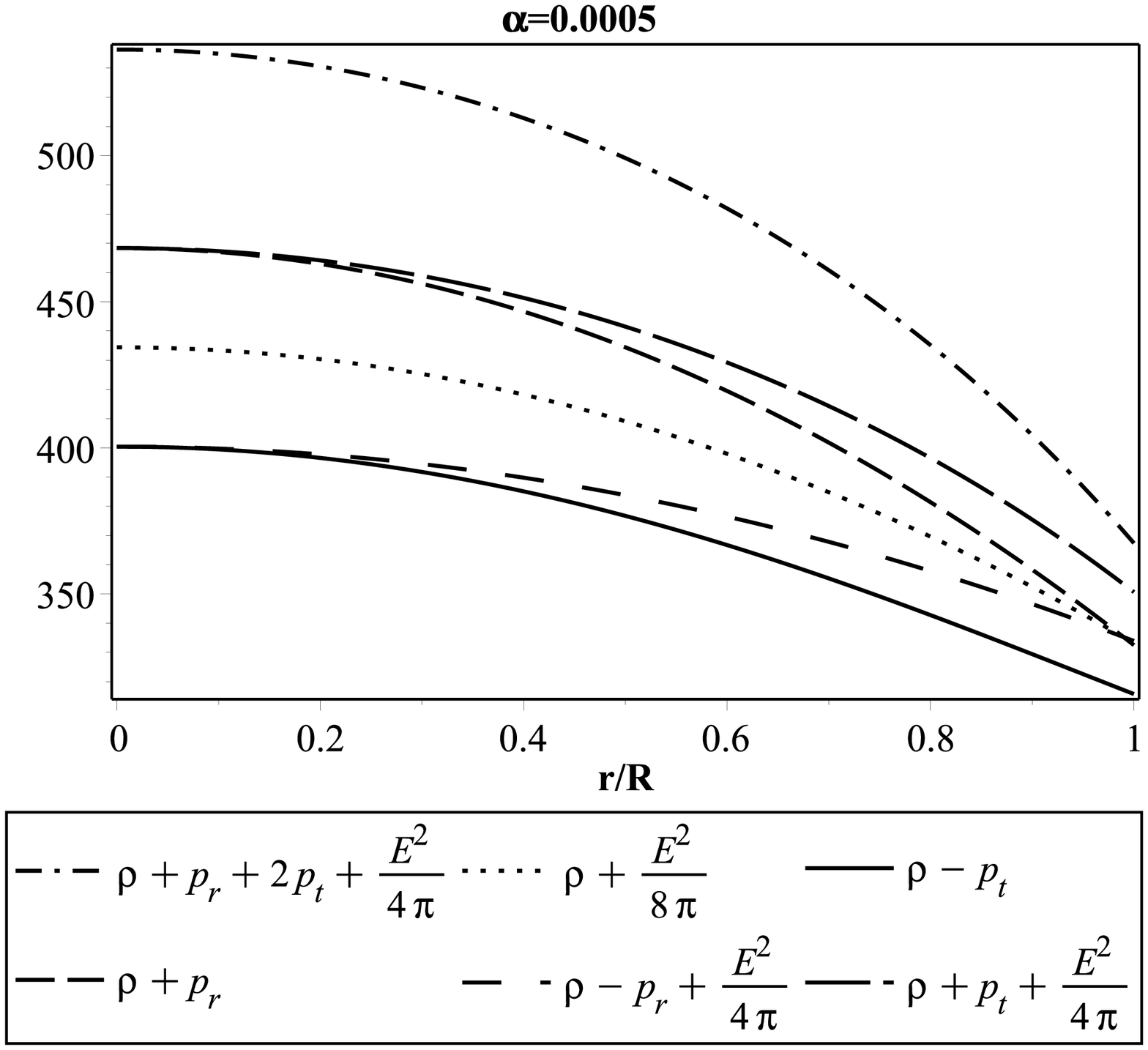}
    \includegraphics[scale=0.2]{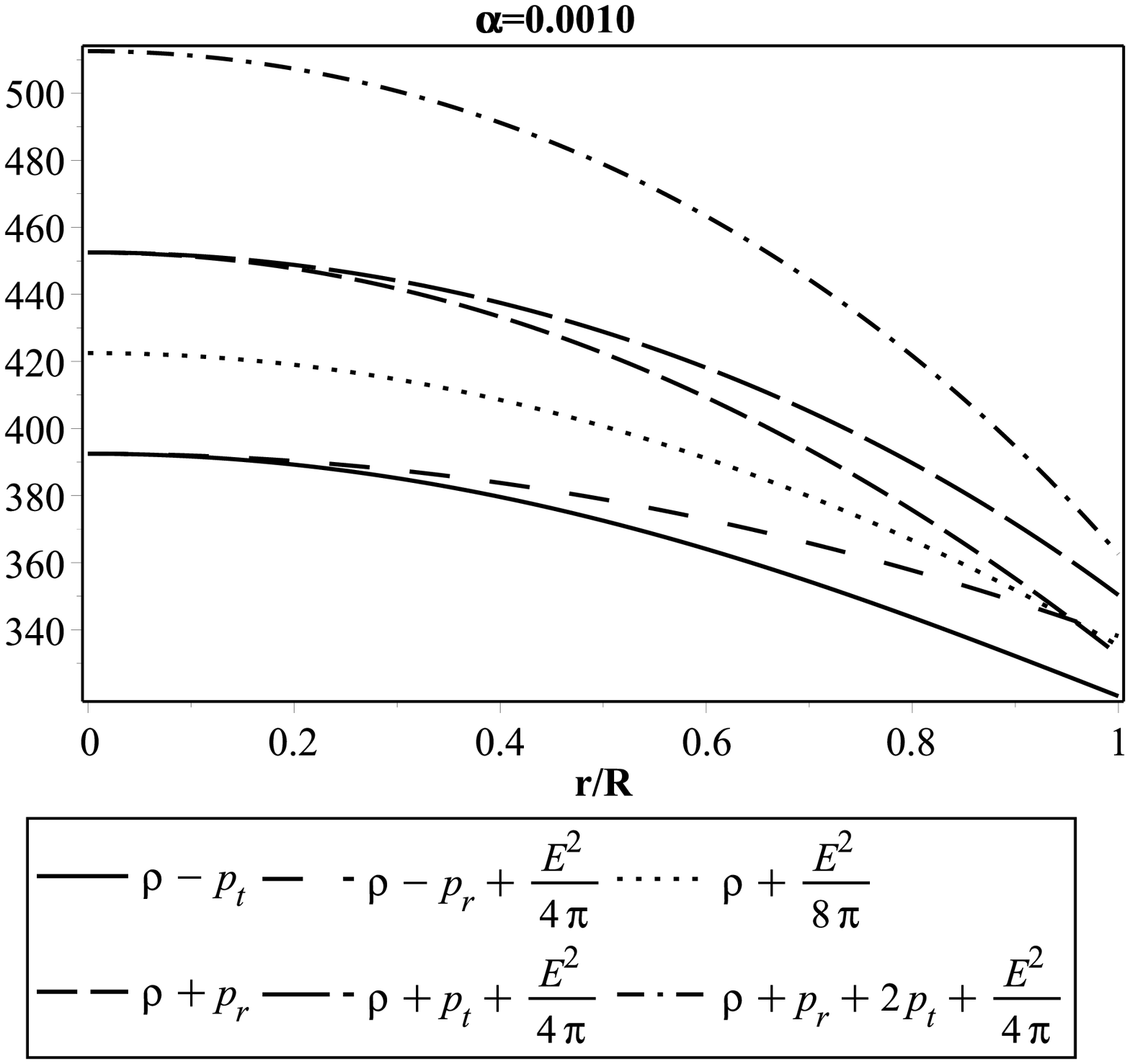}
   \includegraphics[scale=0.2]{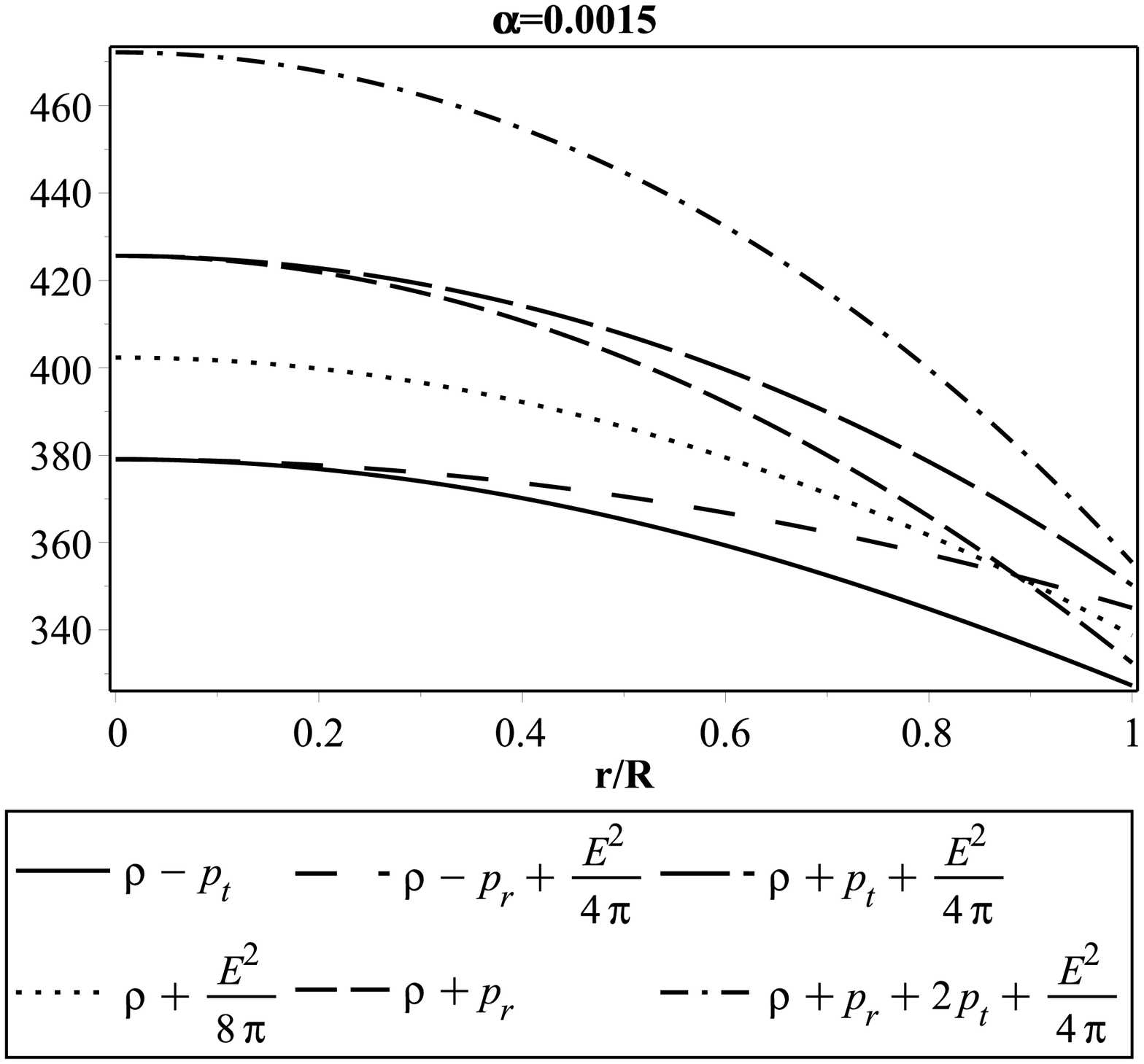}
       \caption{Variation of energy conditions with the radial coordinate $r/R$ for $LMC~X-4$ due to different chosen values of $\alpha$} \label{Fig5}
\end{figure}

We have featured all the inequalities in Fig.~\ref{Fig5} due to the different values of $\alpha$ and Fig.~\ref{Fig5} shows that our system is consistent with all the energy conditions.

\subsection{Mass-radius relation}\label{subsec5.2}
Andr{\'e}asson~\cite{Andreasson2009} predicted the upper bound of the mass-radius ratio for the charged spherically symmetric stellar system, which was generalization of the Buchdahl limit~\cite{Buchdahl1959} that provides upper limit for the allowed mass-radius ratio in the uncharged case. Hence, in the present system the upper bound~\cite{Andreasson2009} is given as
\begin{eqnarray}\label{5.2.1}
{\frac {2M}{R}}\leq {\frac {2}{9{R}^{2}}}\left[3\,{Q}^{2}+2\,{R}^{2}+2\,R\sqrt {3\,{Q}^{2}+{R}^{2}}\right].
\end{eqnarray}

The mass function for our system is provided as follows
\begin{eqnarray}\label{5.2.2}
&\qquad\hspace{-1.2cm} m \left( r \right) = - \frac{{r}^{3}}{2 R^5} \Big[ 3\,{R}^{7}{\alpha}^{2}-3\,{R}^{5}{
\alpha}^{2}{r}^{2}+16\,B\pi \,{R}^{5}\nonumber \\
&\qquad\hspace{1.6cm}-16\,B\pi \,{R}^{3}{r}^{2}  -5\,M{R}^{2}+3\,M{r}^{2} \Big]. 
\end{eqnarray}

\begin{figure}[!htpb]\centering
\includegraphics[scale=0.3]{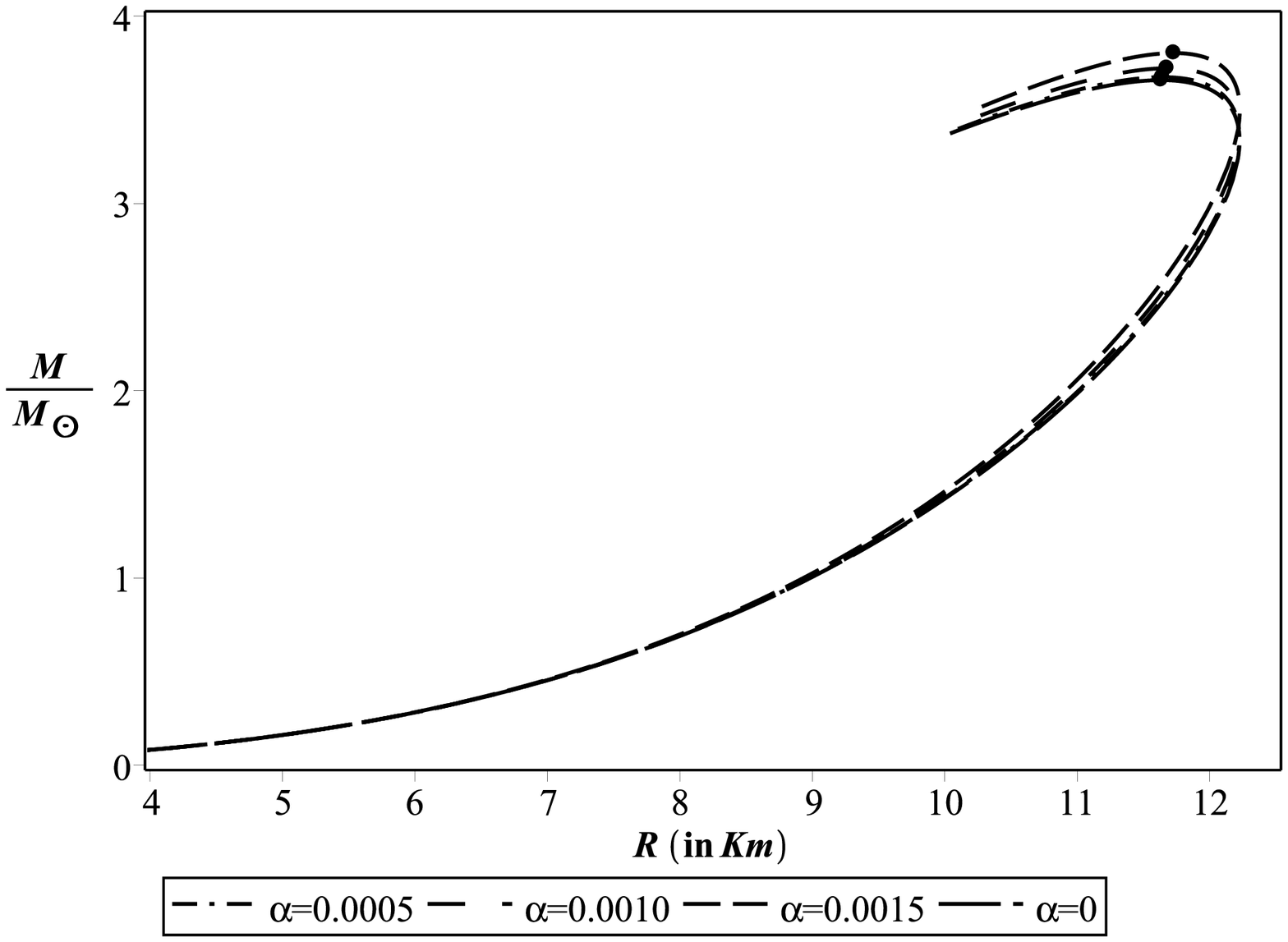}
\includegraphics[scale=0.3]{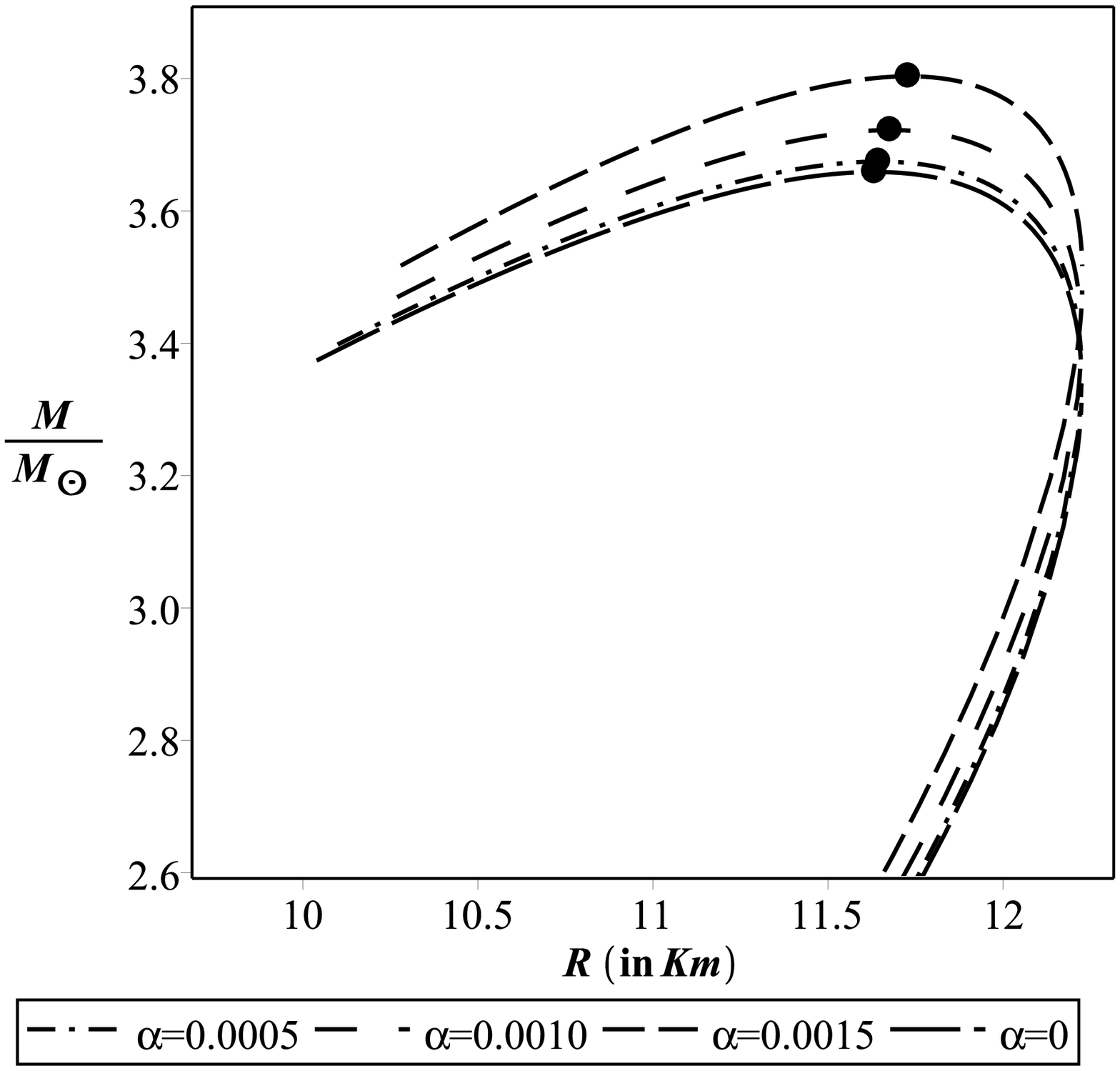}
	\caption{ (i) The upper panel features Mass~$(M/{M_{\odot}})$ vs Radius~($R$~in km) curve for the strange stars due to the different values of $\alpha$ and (ii) the lower panel shows enlarged figure of the $M-R$ curve. The solid circles are representing maximum mass points}  \label{Fig6}
\end{figure}

We have presented variation of the total mass, $M$ (normalized in solar mass, $M_{\odot}$) with respect to the total radius, $R$ due to different parametric values of $\alpha$ in Fig.~\ref{Fig6}, where we chose that the bag constant is $B=83~ MeV/{{fm}^3}$~\cite{Rahaman2014}. We find the maximum mass of the system increases as the value of $\alpha$ increases, which is clearly shown in the lower panel of Fig.~\ref{Fig6}.


\begin{figure}[!htpb]
\centering
\includegraphics[scale=0.2]{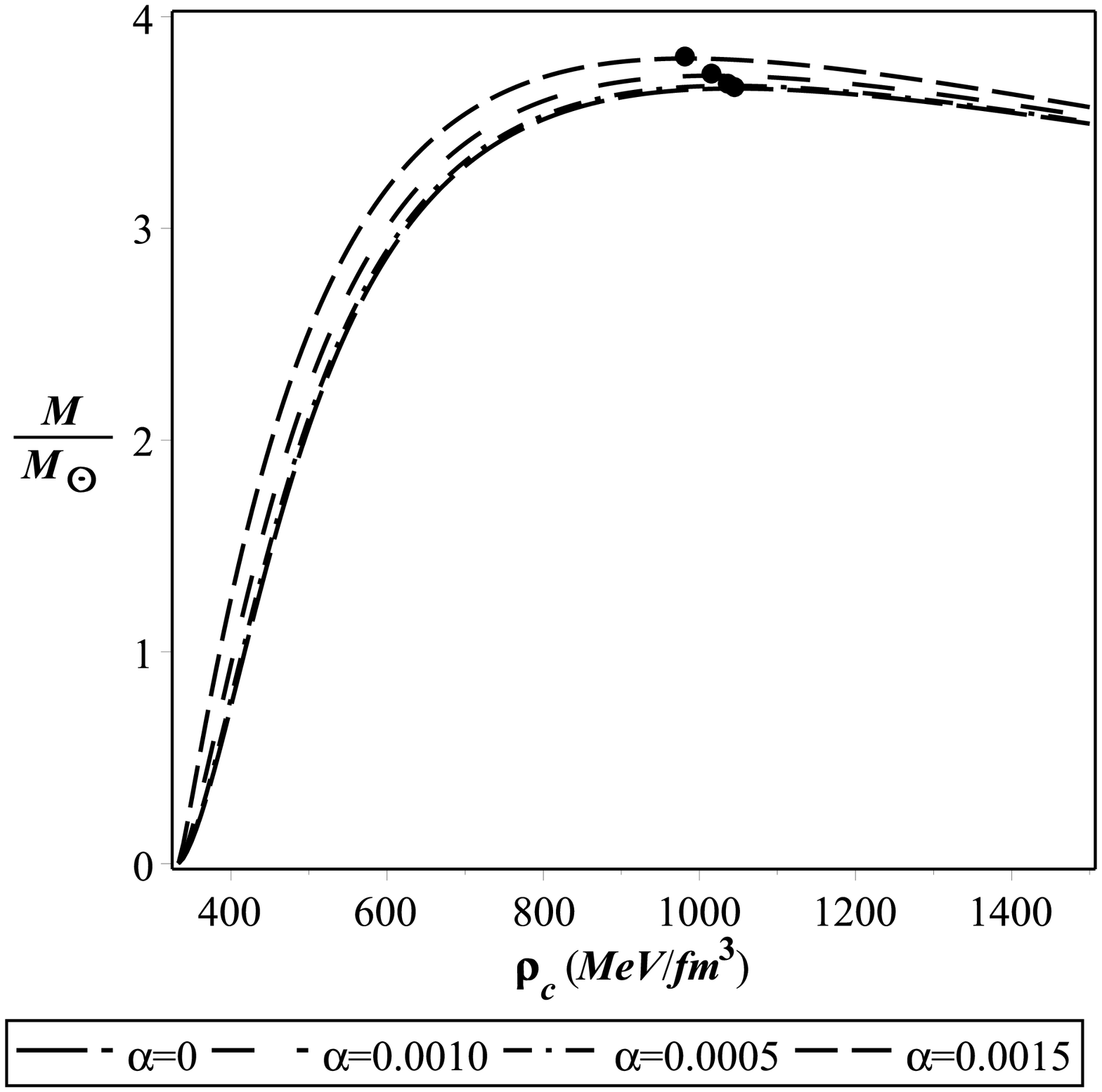}
\includegraphics[scale=0.2]{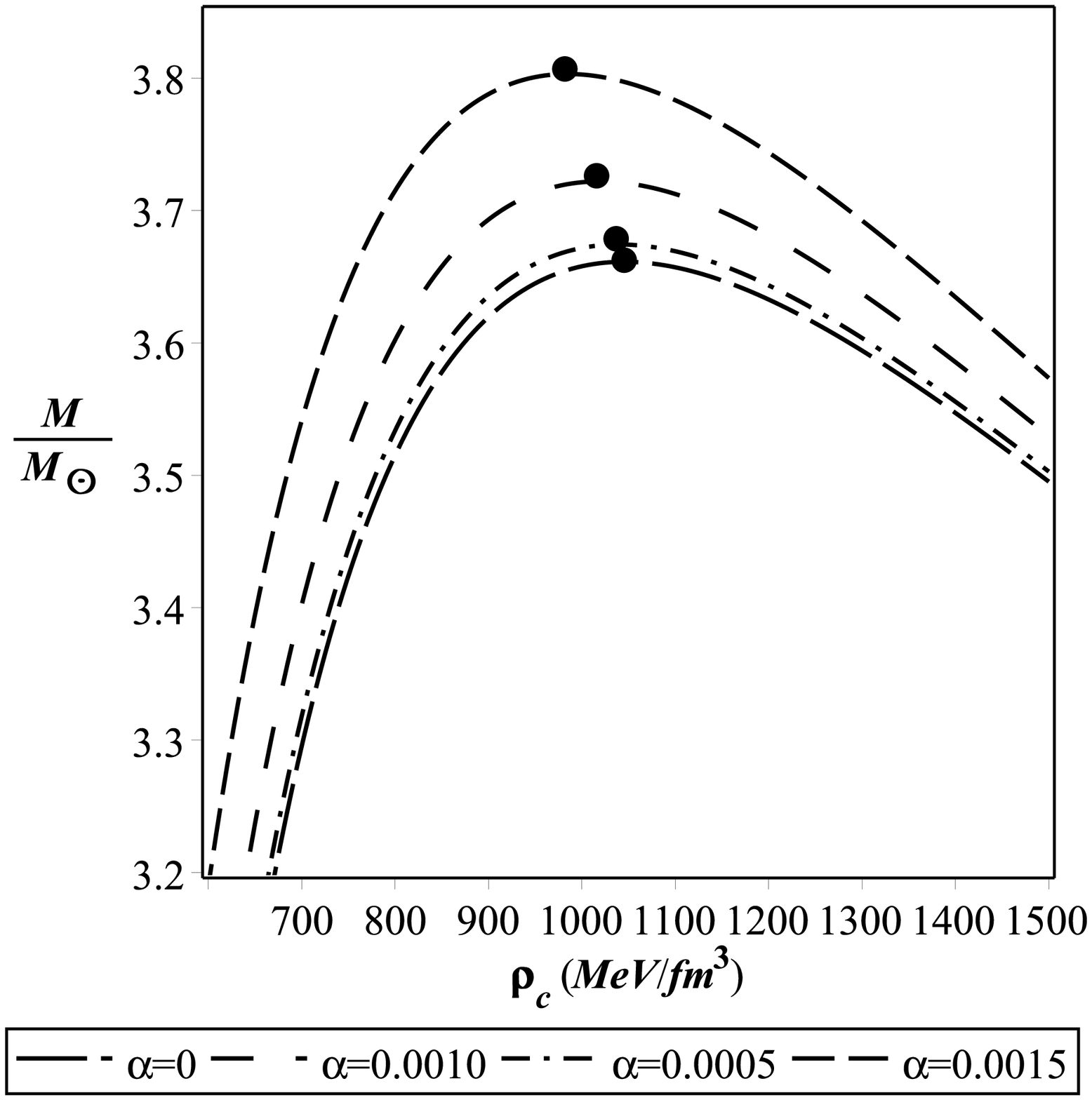}
\includegraphics[scale=0.2]{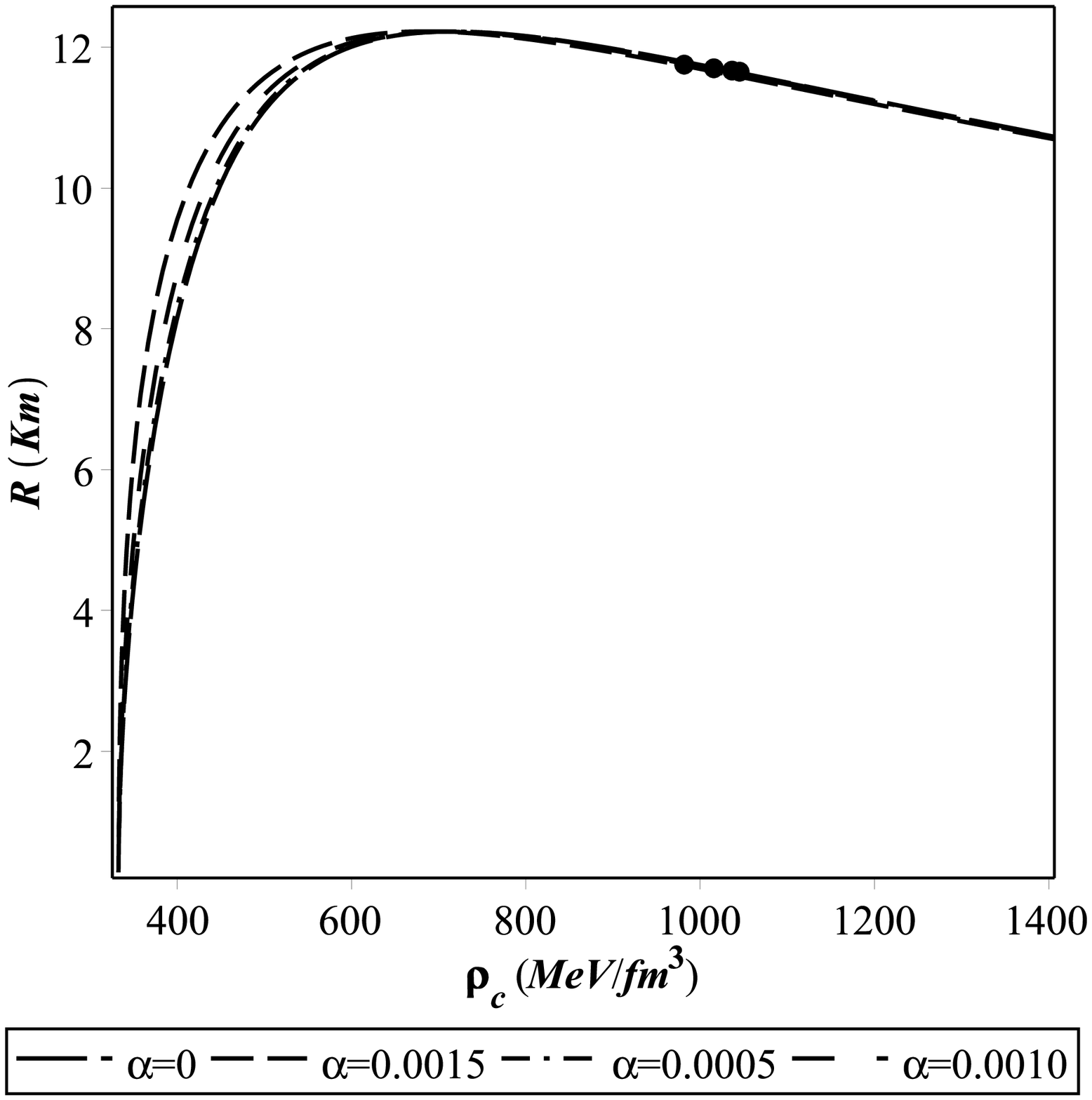}
\includegraphics[scale=0.2]{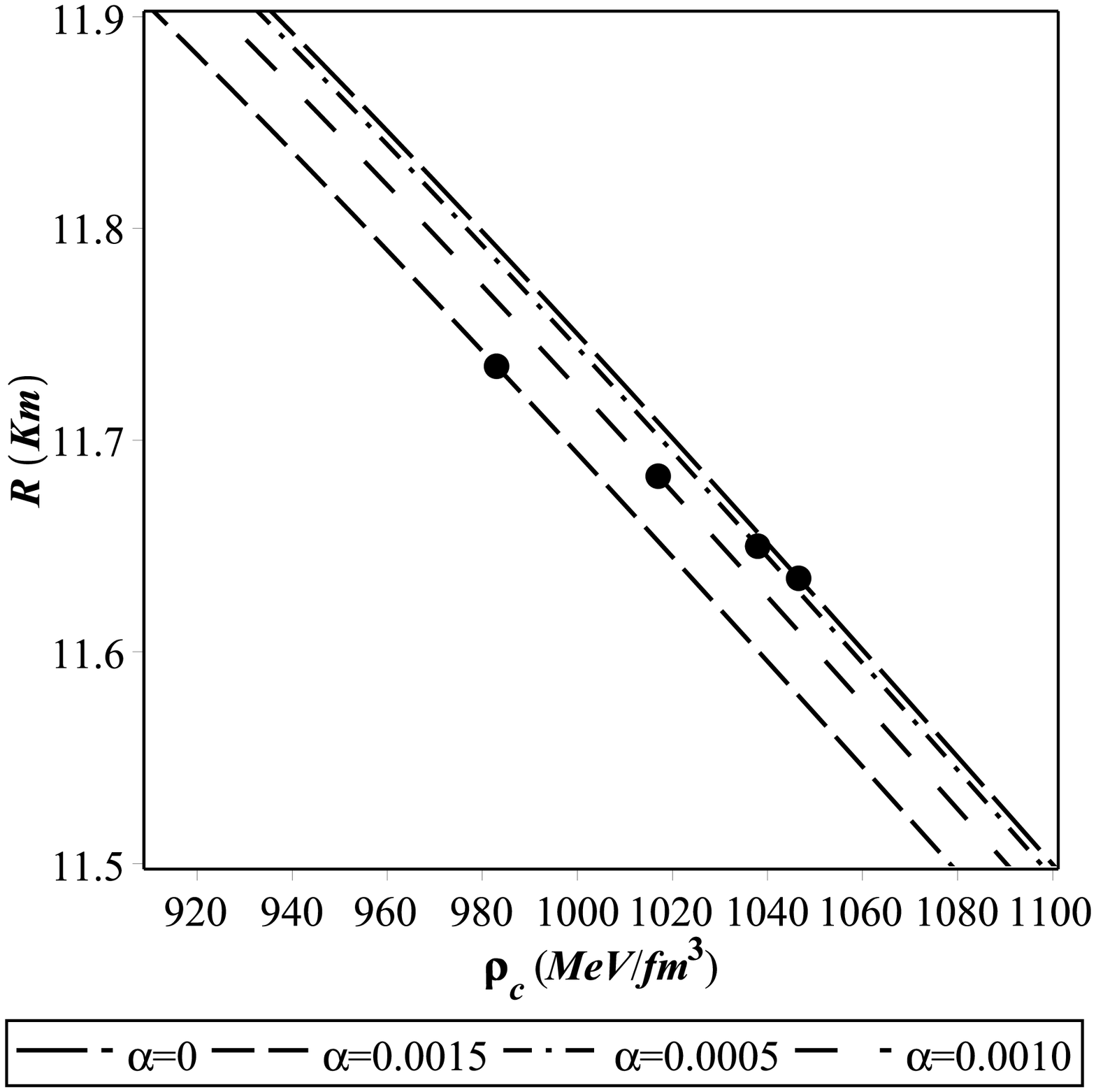}
\caption{Variation of the (i) mass~$M/{M_{\odot}}$ (upper left panel) and (ii) radius~$R$~in km (lower left panel) of the strange stars as a function of the central density $({\rho}_{c})$ are shown. The enlarged versions of the $M/{M_{\odot}}$ vs $({\rho}_{c})$ and $R$ vs $({\rho}_{c})$ curves are shown in the upper right panel and the lower right panel. Here, solid circles are representing maximum mass points}  \label{Fig7}
\end{figure}


We have featured variation of $M$ (normalized in $M_{\odot}$) with respect to the central density ${\rho}_c$ in the left and right upper panel of Fig.~\ref{Fig7}. For $\alpha=0$ the maximum mass, $M_{max}=3.66~{M_{\odot}}$ is achieved for ${\rho}_c=1.866\times {{10}^{15}}~gm/{cm}^3$, whereas for $\alpha=0.0015$ the value of $M_{max}$ increases to $M_{max}=3.81~{M_{\odot}}$ and the value of the corresponding central density decreases to ${\rho}_c=1.753\times {{10}^{15}}~gm/{cm}^3$. The left and right lower panel of Fig.~\ref{Fig7} show the variation of $R$ with respect to ${\rho}_c$. We find, $R_{Mmax}$, the radius corresponding to $M_{max}$ decreases from $11.734~km$ to $11.634~km$ as the value of $\alpha$ decreases from $\alpha=0.0015$ to $\alpha=0$, respectively.

\subsection{Compactification factor and redshift}\label{subsec5.3}
The compactification factor for our system is defined as
\begin{eqnarray}\label{5.3.1}
&\qquad\hspace{-1.2cm} u=\frac{m(r)}{r} = - \frac{{r}^{2}}{2 R^5} \Big[ 3\,{R}^{7}{\alpha}^{2}-3\,{R}^{5}{
\alpha}^{2}{r}^{2}+16\,B\pi \,{R}^{5}\nonumber \\
&\qquad\hspace{1.6cm}-16\,B\pi \,{R}^{3}{r}^{2}  -5\,M{R}^{2}+3\,M{r}^{2} \Big]. 
\end{eqnarray}

Hence, the surface redshift, $Z_s$ corresponding to  the compactification factor $u$ is given as
\begin{eqnarray}\label{5.3.2}
Z_{{s}}={\frac {1}{\sqrt {1-2\,u \left( R \right) }}}-1={\frac {\sqrt {R}}{\sqrt {{R}^{5}{\alpha}^{2}-2\,M+R}}}.
\end{eqnarray}


\begin{figure}[!htpb]
\centering
\includegraphics[scale=0.3]{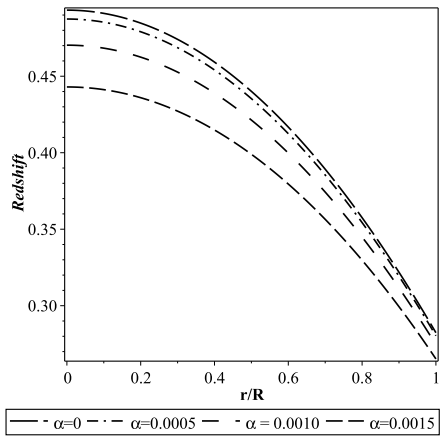}
\caption{Variation of the redshift function with respect to the radial coordinate $r/R$ for the strange star $LMC~X-4$ }  \label{Fig8a}
\end{figure}


The variation of the redshift function, $Z(r)$ with respect to the radial coordinate $r/R$ is presented in Fig.~\ref{Fig8a}. Clearly, in a spherically symmetric anisotropic charged stellar system as the value of $\alpha$ increases the values of the surface redshift gradually decreases.

\subsection{The stability of the system}\label{subsec5.4}
To examine stability of our system we will study i) Generalized TOV equation and ii) Herrera cracking concept as follows

\subsubsection{Generalized TOV equation}\label{subsubsec5.4.1}
\begin{figure}[!htpb]
\centering
\includegraphics[scale=0.2]{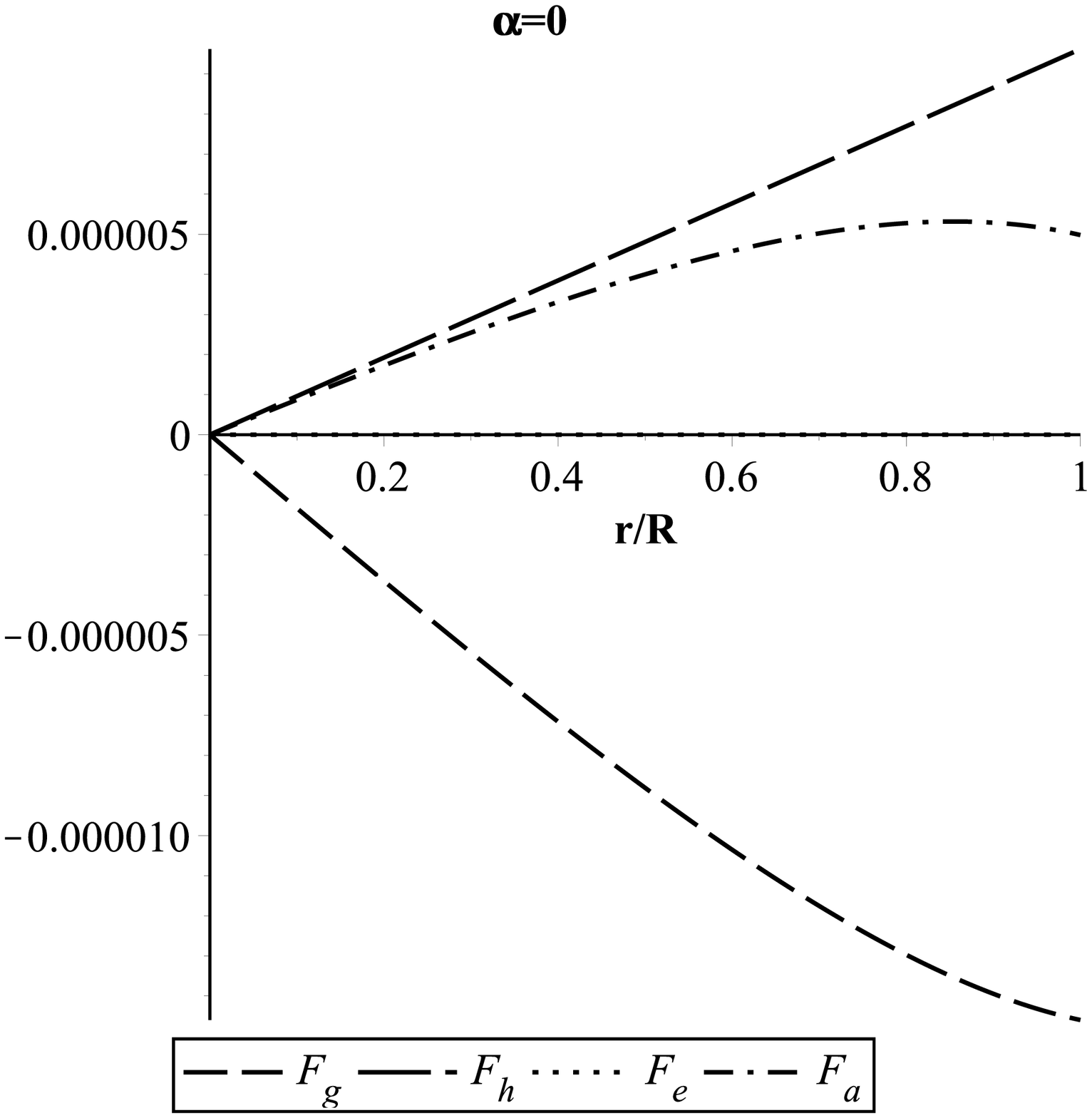}
\includegraphics[scale=0.2]{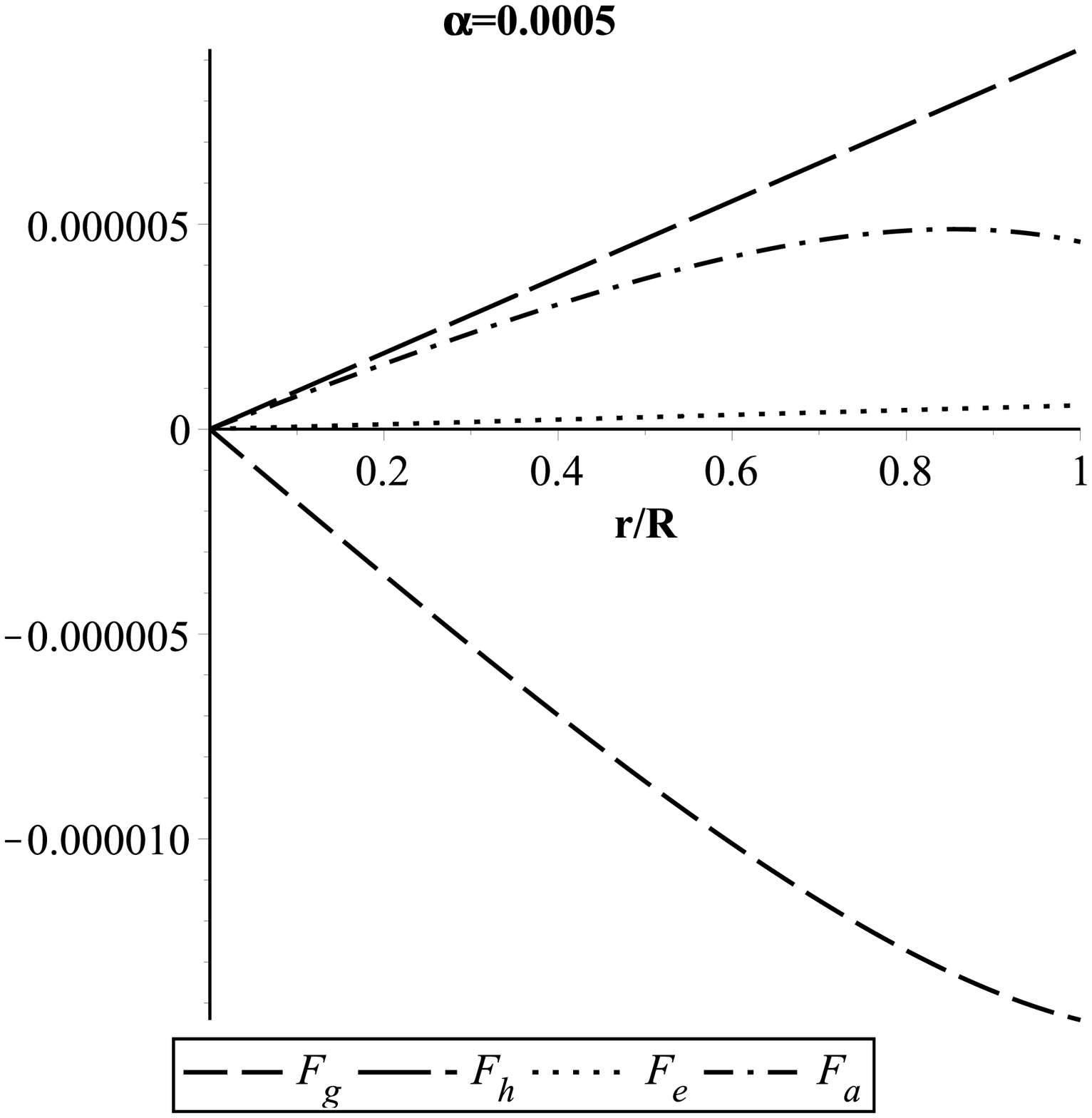}
\includegraphics[scale=0.2]{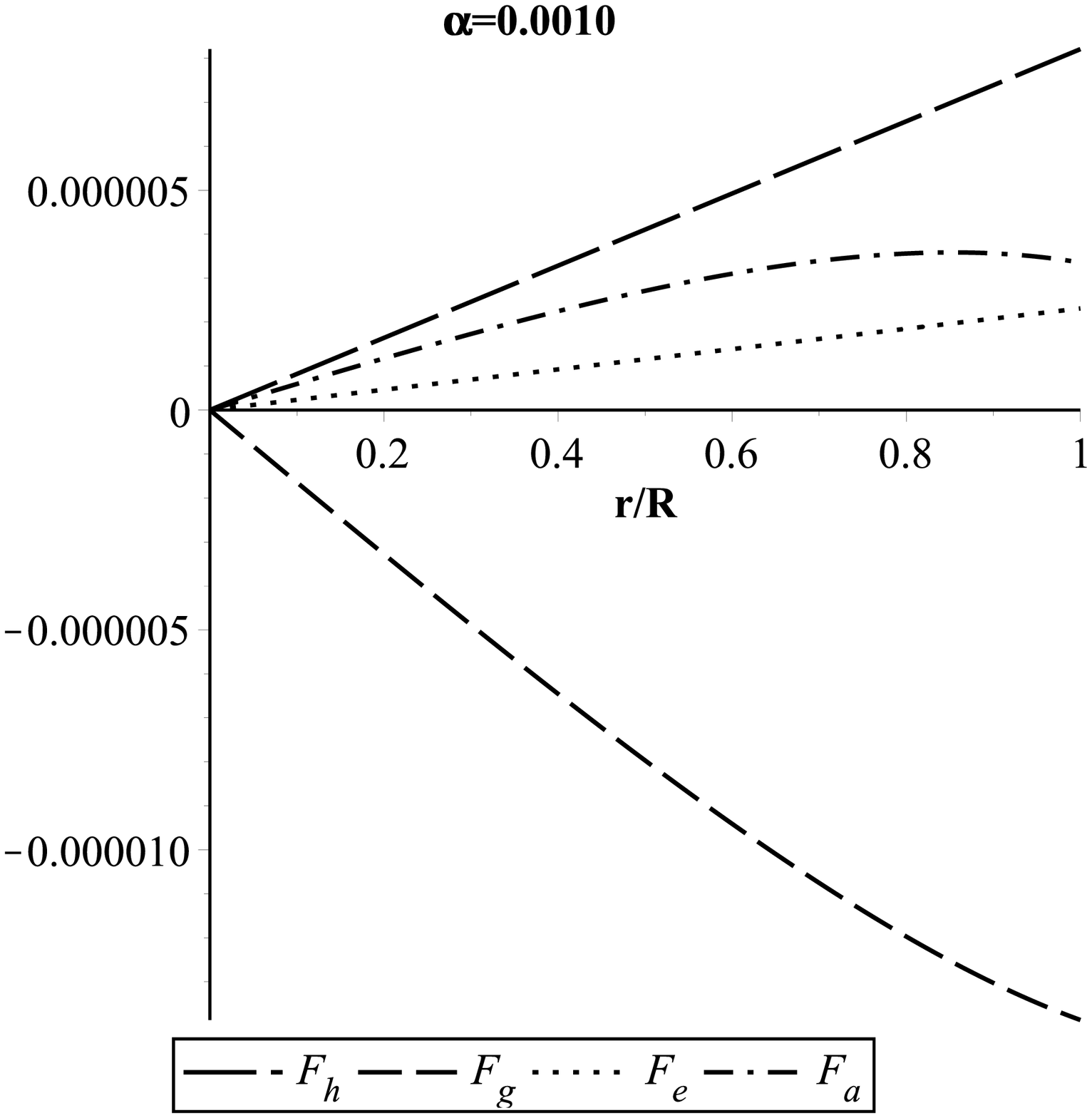}
\includegraphics[scale=0.2]{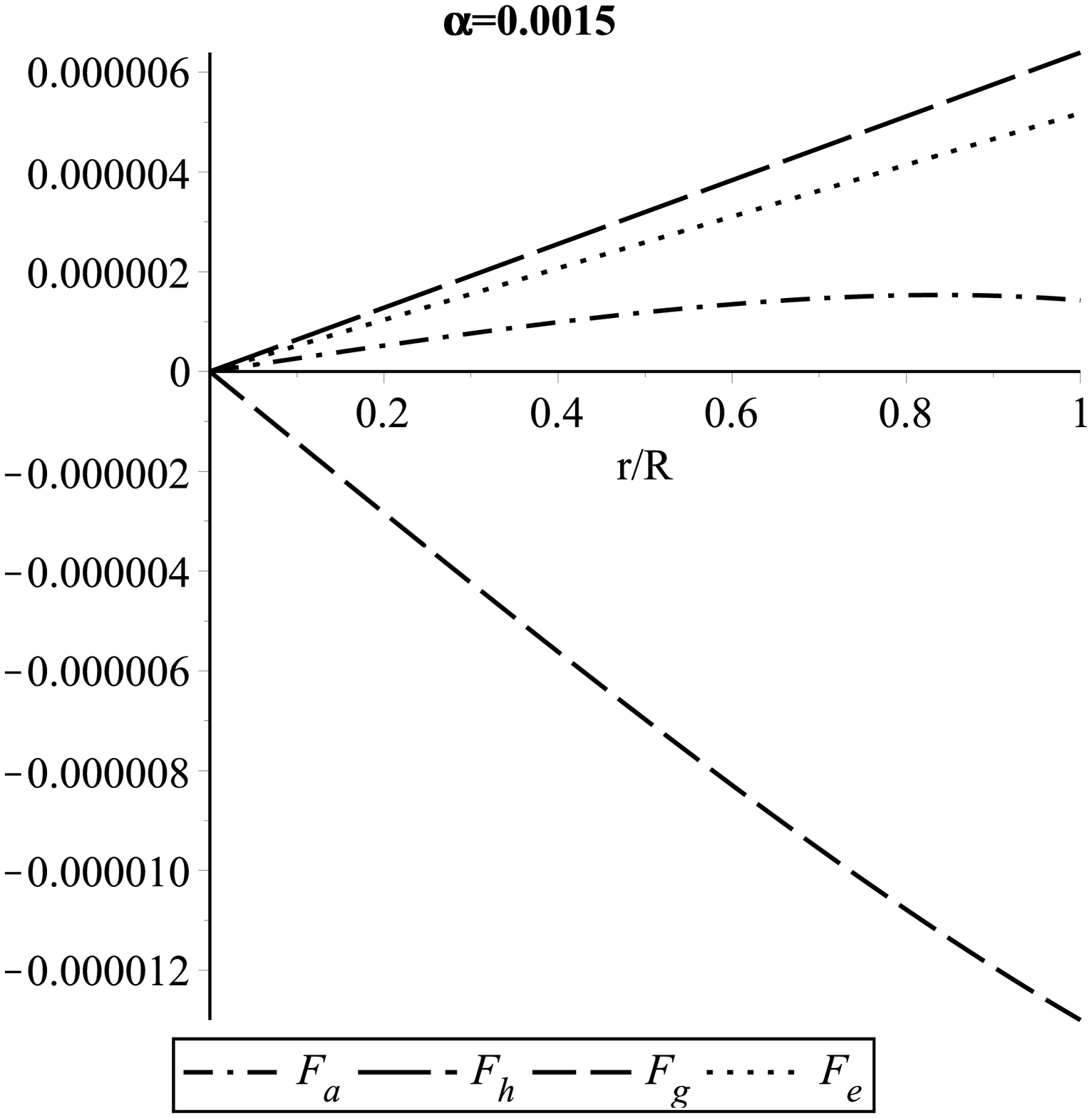}
\caption{Variation of the different forces as a function of the radial coordinate $r/R$ for the strange stars $LMC~X-4$}  \label{Fig8}
\end{figure}

The generalized form of the TOV equation in the present anisotropic charged system reads
\begin{eqnarray}\label{5.4.1.1}
-\frac{M_g(\rho +p_r)}{r^{2}} e^{\frac{\lambda- \nu }{2}}-\frac{d p_r}{dr}+ \sigma \frac{q}{r^2} {e^{\frac{\lambda}{2}}} + \frac{2}{r} (p_t-p_r)=0,\nonumber\\
\end{eqnarray}
where $M_g$ denotes the effective gravitational mass and given as follows
\begin{eqnarray}\label{5.4.1.2}
M_g (r)=\frac{1}{2} {r^2}{e^{\frac{\nu-\lambda}{2}}}{{\nu}^{\prime}}.
\end{eqnarray}
Eq.~(\ref{5.4.1.1}) features that the system is completely stable under the equilibrium of the different forces, i.e., $F_g+F_h+F_e+F_a=0$, where $F_g$, $F_h$, $F_e$ and $F_a$ represents gravitational, hydrodynamic, electric and anisotropic force, respectively. We have presented variation of the different forces with respect to the radial coordinate $r/R$ due to different values of $\alpha$ in Fig.~\ref{Fig8}. The figure features that the attractive gravitational force $F_g$, which acts toward the inward direction along the system is counterbalanced by the combined effects of the forces $F_h$, $F_e$ and $F_a$.


\subsubsection{Herrera cracking concept}\label{subsubsec5.4.2}
To examine stability of the system in terms sound speeds the systems have to be consistent with the i) causality condition and ii) Herrera cracking concept. To be consistent with the causality condition, the square of the radial ($v^2_{sr}$) and tangential ($v^2_{st}$) sound speeds should satisfy the inequalities $0 \leq v^2_{sr} \leq 1$ and $0 \leq v^2_{st} \leq 1$ simultaneously. According to the concept of Herrera's cracking~~\cite{Herrera1992,Abreu2007} for a potentially stable region $v^2_{sr}$ should be greater than $v^2_{st}$ and the difference of the square of the sound speeds should maintain it's sign same through out that region, i.e.,  $|{v^2_{st}}- {v^2_{sr}}|\leq 1$. The square of sound speeds are defined as
\begin{eqnarray}\label{5.4.2.1}
 v^2_{sr}=\frac{p_r}{\rho},\\ \label{5.4.2.2}
 v^2_{st}=\frac{p_t}{\rho}.
\end{eqnarray} 


\begin{figure}[!htpb]
\centering
\includegraphics[scale=0.3]{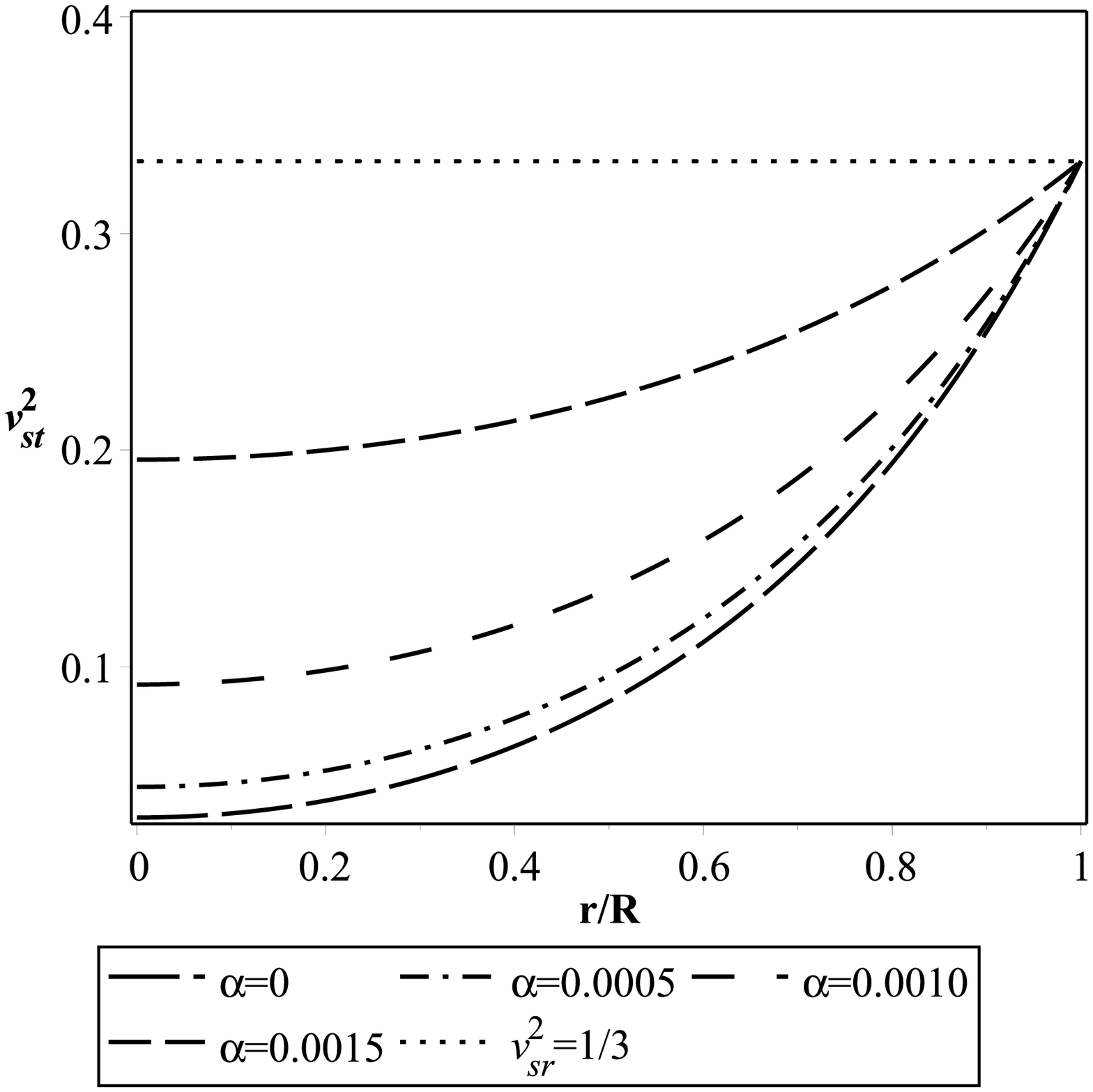}
\includegraphics[scale=0.3]{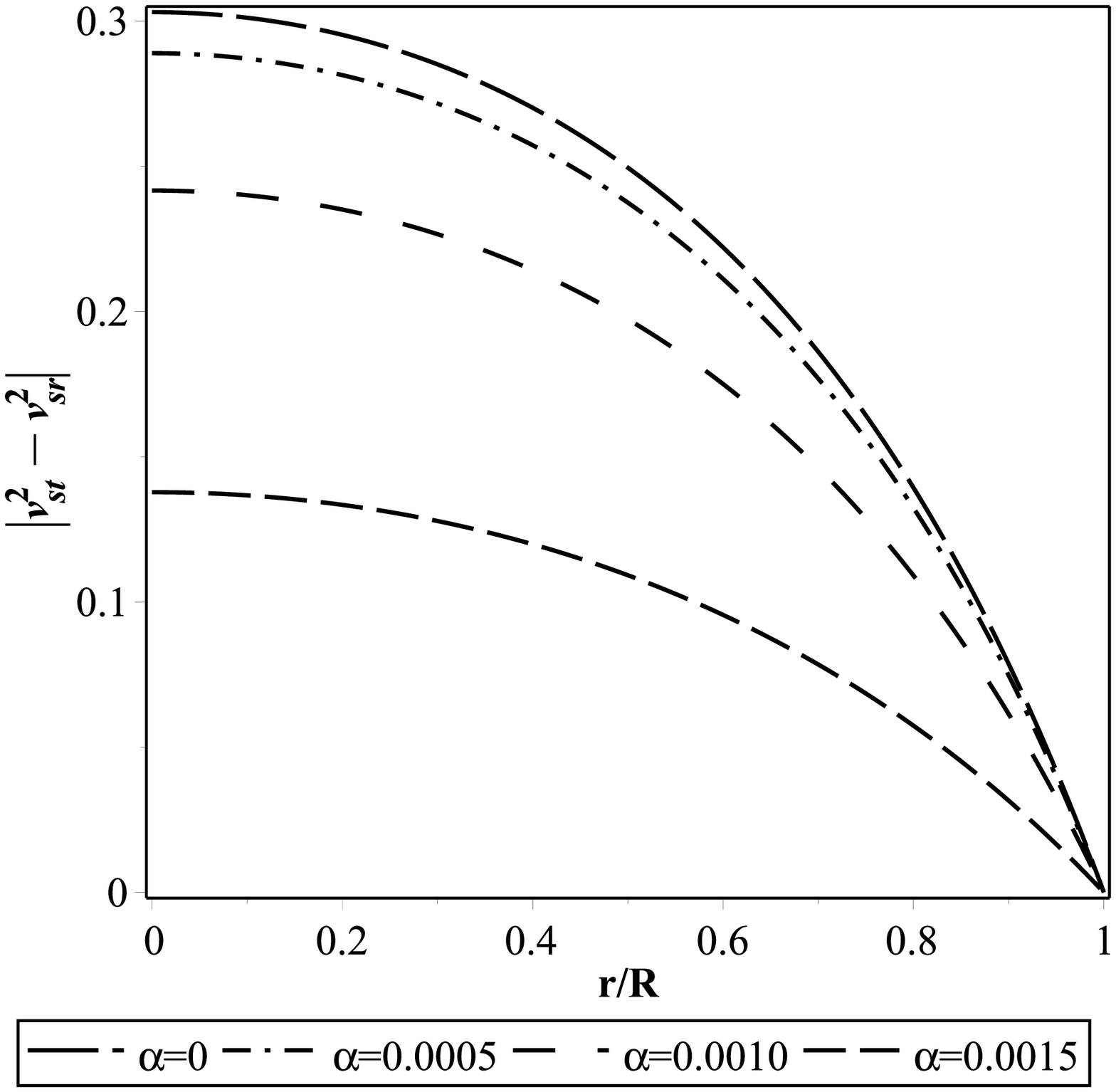}
\caption{Variation of (i) $v^2_{sr}$ and $v^2_{st}$ (upper panel) and (ii) $|{v^2_{st}}- {v^2_{sr}}|\leq 1$ (lower panel) as a function of the radial coordinate}  \label{Fig9}
\end{figure}


In Fig.~\ref{Fig9} we have shown the variation of the square of the sound speeds (upper panel) and $|{v^2_{st}}- {v^2_{sr}}|$ (lower panel) with respect to the radial coordinate $r/R$ due to different parametric values of $\alpha$. The figure clearly features that both our system satisfies both the causality condition and Herrera cracking concept. Hence, our system is completely stable.

\section{Discussion and conclusion}\label{sec6}
In this literature we have presented a detailed study on the effect of the electrical charge on the spherically symmetric anisotropic stellar system, which is made of SQM and governed by the MIT bag EOS. Assuming a simplified form of the electrical charge distribution given as $q\left(r\right)=Q\left({r}/{R} \right)^3\equiv \alpha\,{r^3}$, we have obtained exact solutions for the Maxwell-Einstein field equations. Further, using the exterior Reissner-Nordstr{\"o}m metric we have presented expressions for the different physical parameters in Eqs.~(\ref{4.1})-(\ref{4.6}). We have presented the obtained solutions and studied their physical validity in terms of the star $LMC~X-4$ of mass $1.29~M_{\odot}$, by considering it as the representative of strange star candidates. Throughout the study we have considered bag constant as $B=83~ MeV/{{fm}^3}$ and the chosen parametric values of $\alpha$~(in~${km}^{-2}$) as $0$, $0.0005$, $0.0010$ and $0.0015$.

 The profile of the metric potentials $\left( {e^{\nu}}, {e^{\lambda}} \right)$ are shown in Fig.~\ref{Fig1}, which shows that at the center both the metric potentials are finite. It confirms that our system is free from any sort of singularities, i.e., physical or geometrical singularities. The variations of $\rho$, $p_r$ and $p_t$ are shown in the upper, middle and lower panel in Fig.~\ref{Fig2}, respectively. We find that density and pressure functions are maximum at the surface and decrease monotonically through out the system to reach the minimum value at the surface and confirms regularity of the achieved solutions. We have predicted different values of the central density, ${\rho}_c$ and central pressure ${p}_c$ for the different strange star candidates in Table~\ref{Table 1}. We find that the densities and radial pressures of the different strange stars are in the order of ${10}^{14}~gm/{cm}^3$ and ${10}^{34}~dyne/{cm}^2$, respectively. Due to the strange star candidates as mentioned in Table~\ref{Table 1} we find density is much higher than the normal nuclear density ${\rho}_{normal}=2.3\times {10}^{14}~{gm}/{cm}^3$, which confirms that the stars are made of SQM. The variation of the anisotropic stress for the different values of $\alpha$ is shown in Fig.~\ref{Fig3} and it confirms the prediction by Deb et al.~\cite{Deb2017} that for an anisotropic strange star the anisotropic stress should be maximum at the surface.


\begin{table}[htbp!]
\centering
\caption{Numerical values of physical parameters for the different strange stars for $\alpha=0.0010~{km}^{-2}$ and $B=83~ MeV/{{fm}^3}$~\cite{Rahaman2014} } \label{Table 1}
\rotatebox{90}{
\begin{tabular}{ ccccccccccccccccccccccccccc}
\hhline{=========}
Strange   & Observed  & Predicted  & ${\rho}_c$ & ${p}_c$ & $Q$ & $E$ & $\frac{2M}{R}$ & $Z_s$ \\ 
 Stars & Mass $\left(  M_{\odot} \right)$ & Radius $\left( km \right)$ & $\left( gm/{{cm}^3} \right)$ & $\left( dyne/{{cm}^2} \right)$ & $\left( Coulomb \right)$ & $\left(V/cm\right)$ &  &  \\
\hline 
$PSR~J1614-2230$ & $1.97 \pm 0.04$ \cite{demorest} & $10.927\pm 0.060$ & $8.439 \times {{10}^{14}}$ & $7.526 \times {{10}^{34}}$ & $1.514 \times {{10}^{20}}$ & $1.140 \times {{10}^{22}}$ & 0.531 & 0.460 \\ 
\
$Vela~X-1$ & $1.77 \pm 0.08$\cite{dey2013} & $10.608 \pm 0.135$ & $8.150 \times {{10}^{14}}$ & $6.659 \times {{10}^{34}}$ & $1.385 \times {{10}^{20}}$ & $1.107 \times {{10}^{22}}$ & 0.492 & 0.403 \\   
\
$4U~1608-52$ & $1.74 \pm 0.14$\cite{guver2010a} & $10.558 \pm 0.240$ & $8.104 \times {{10}^{14}}$ & $6.521 \times {{10}^{34}}$ & $1.366 \times {{10}^{20}}$ & $1.102 \times {{10}^{22}}$ & 0.486 & 0.395 \\  
\
$PSR~J1903 + 327$ & $1.667 \pm 0.021$\cite{dey2013} & $10.430 \pm 0.037$ & $8.007 \times {{10}^{14}}$ & $6.232 \times {{10}^{34}}$ & $1.317 \times {{10}^{20}}$ & $1.088 \times {{10}^{22}}$ & 0.472 & 0.376 \\  
\
$4U~1820-30$ & $1.58 \pm 0.06$\cite{guver2010b} & $10.271 \pm 0.112$ & $7.893 \times {{10}^{14}}$ & $5.889 \times {{10}^{34}}$ & $1.257 \times {{10}^{20}}$ & $1.072 \times {{10}^{22}}$ & 0.454 & 0.353 \\  
\
$Cen~X-3$ & $1.49 \pm 0.08$\cite{dey2013} & $10.098 \pm 0.158$ & $7.777 \times {{10}^{14}}$ & $5.543 \times {{10}^{34}}$ & $1.195 \times {{10}^{20}}$ & $1.054 \times {{10}^{22}}$ & 0.435 & 0.330 \\  
\
$EXO~1785-248$ & $1.3 \pm 0.2$\cite{ozel2009} & $9.701 \pm 0.446$ & $7.541 \times {{10}^{14}}$ & $4.836 \times {{10}^{34}}$ & $1.059 \times {{10}^{20}}$ & $1.012 \times {{10}^{22}}$ & 0.395 & 0.286 \\  
\
$LMC~X - 4$ & $1.29 \pm 0.05$\cite{dey2013} & $9.678 \pm 0.111$ & $7.533 \times {{10}^{14}}$ & $4.812 \times {{10}^{34}}$ & $1.052 \times {{10}^{20}}$ & $1.010 \times {{10}^{22}}$ & 0.393 & 0.284 \\  
\
$SMC~X - 1$ & $1.04 \pm 0.09$\cite{dey2013} & $9.070 \pm 0.240$ & $7.241\times {{10}^{14}}$ & $3.937 \times {{10}^{34}}$ & $0.866 \times {{10}^{20}}$ & $9.465 \times {{10}^{21}}$ & 0.338 & 0.229 \\   
\
$\hspace{-0.2cm} SAX~J1808.4-3658$ & $0.9 \pm 0.3$\cite{dey2013} & $8.676 \pm 0.914$ & $7.089\times {{10}^{14}}$ & $3.480 \times {{10}^{34}}$ & $0.758 \times {{10}^{20}}$ & $9.053 \times {{10}^{21}}$ & 0.306 & 0.200 \\  
\
$4U~1538-52$ & $0.87 \pm 0.07$\cite{dey2013} & $8.586 \pm 0.214$ & $7.053\times {{10}^{14}}$ & $3.374 \times {{10}^{34}}$ & $0.734 \times {{10}^{20}}$ & $8.960 \times {{10}^{21}}$ & 0.298 & 0.194 \\  
\
$HER~X-1$ & $0.85 \pm 0.15$\cite{dey2013} & $8.524 \pm 0.469$ & $7.034\times {{10}^{14}}$ & $3.316 \times {{10}^{34}}$ & $0.719 \times {{10}^{20}}$ & $8.895 \times {{10}^{21}}$ & 0.294 & 0.190 \\ 
\hhline{=========} 
\end{tabular}  }
  \end{table}


 The profile of the electrical charge $q(r)$ and electrical energy density ${E^2}(r)/8\pi$ is featured in the upper and lower panel in Fig.~\ref{Fig4}, respectively. We find that the total charge $Q$ and the associated electric field $E$ are in the order of ${10}^{20}~C$ and ${{10}^{21}}-{{10}^{22}}~V/cm$, respectively. Our study clearly reveals that the electric charge has a significant effect on the different physical parameters and the stability of the anisotropic spherically symmetric system. Both Fig.~\ref{Fig2} and Table {\ref{Table 2} shows that as the charge increases the density and pressures of the stellar system decreases gradually. Interestingly, Fig.~\ref{Fig3} features that the effect of anisotropy on the stellar system is maximum when the system is neutral. However, the anisotropic stress of the system decreases consequently with the increasing effect of the charge.


\begin{figure}[!htp]\centering
	\includegraphics[scale=.3]{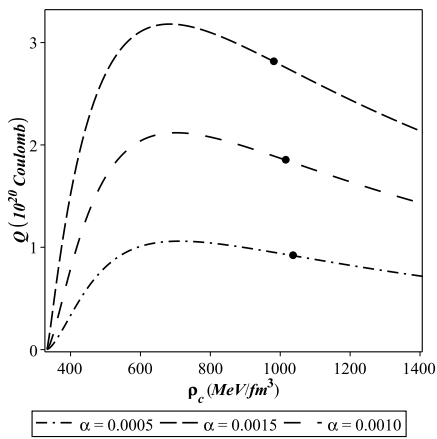}
			\caption{Variation of $Q$ in coulomb as a function of the central density ${\rho}_c$ for the strange star $LMC~X-4$}\label{Fig10}
\end{figure}



\begin{table}[htbp!]
  \centering
    \caption{Numerical values of the different physical parameters for the different strange stars for $LMC~X-4$ due to different values of $\alpha$ with $B=83~ MeV/{{fm}^3}$~\cite{Rahaman2014} } \label{Table 2}
     \rotatebox{90}{
  \begin{tabular}{cccccccccccccccccccccccccc}
\hhline{========} 
Values & Predicted & ${{\rho}_c}$ & ${{p}_c}$ & $Q$ & $E$ & $\frac{2M}{R}$ & $Z_s$ \\ 
 of $\alpha$ &  Radius $(km)$ & $\left( gm/{cm}^3\right)$ & $\left( dyne/{cm}^2\right)$ & $\left( Coulomb\right)$ & $\left(V/cm\right)$ &  &  \\ 
\hline
$0$ & $9.711$ & $7.814 \times {{10}^{14}}$ & $5.652 \times {{10}^{34}}$ & $0$ & $0$ &  $0.3919$ & $0.2824$ \\ 
\
$0.0005$ & $9.702$ & $7.747 \times {{10}^{14}}$ & $5.451 \times {{10}^{34}}$ & $0.530 \times {{10}^{20}}$ & $5.062 \times {{10}^{21}}$ & $0.3922$ & $0.2827$ \\ 
\ 
$0.0010$ & $9.678$ & $7.533 \times {{10}^{14}}$ & $4.812 \times {{10}^{34}}$ & $1.052 \times {{10}^{20}}$ & $1.010 \times {{10}^{22}}$ & $0.3932$ & $0.2837$ \\ 
\
$0.0015$ & $9.641$ & $7.171 \times {{10}^{14}}$ & $3.728 \times {{10}^{34}}$ & $1.560 \times {{10}^{20}}$ & $1.509 \times {{10}^{22}}$ & $0.3947$ & $0.2853$ \\ 
\hhline{========}  
\end{tabular} }
   \end{table}


 We perform different physical tests, viz., energy conditions, mass-radius relation, generalized TOV equation and Herrera cracking concept, etc. In Fig.~\ref{Fig5} we have shown that our system is consistent with all the energy conditions. We have featured variation of $M$~(normalized in $M_{\odot}$) with respect to $R$ for the different values of $\alpha$ in Fig.~\ref{Fig6}. The solid circles in Fig.~\ref{Fig6} denotes the maximum mass points due to different values of $\alpha$. We found as the charge increases both $M_{max}$ and $R_{Mmax}$ increases gradually. For $\alpha=0.0015$ the values of $M_{max}$ and $R_{Mmax}$ increases $4.1 \%$ and $0.86 \%$, respectively, than the uncharged case. In the upper and lower panel in Fig.~\ref{Fig7} we have presented variation of $M$ and $R$ with respect to ${\rho}_c$, respectively. For $\alpha=0.0015~{km}^{-2}$ the maximum mass point is achieved for ${\rho}_c=7.024\,{{\rho}_{nuclear}}$, which is $6.07\%$ lower than the value of ${\rho}_c$ as in uncharged case. We have also presented variation of the total charge $Q$ with respect to ${\rho}_c$ due to different values of $\alpha$ in Fig.~\ref{Fig10}. The figure reveals that as the value of $\alpha$ increases the total charge ($Q_{max}$) corresponding to $M_{max}$ is achieved for the lower value of ${\rho}_c$. The variation of the redshift function with respect to $r/R$ is shown in Fig.~\ref{Fig8a}. To examine stability of the system we have studied Generalized TOV equation which predicts that for our system sum of the forces $F_g$, $F_a$, $F_e$~and~$F_h$ is zero and variation of the forces due to different values of $\alpha$ is shown in Fig.~{\ref{Fig8}}. Further, Fig.~\ref{Fig9} features that our system is consistent with the Herrera cracking concept by satisfying all the inequalities simultaneously given as $0 \leq v^2_{sr} \leq 1$, $0 \leq v^2_{st} \leq 1$ and $|{v^2_{st}}- {v^2_{sr}}|\leq 1$. 

In Table~\ref{Table 1} we have predicted a detailed data sheet of the different physical parameters for the different strange star candidates due to $\alpha=0.0010~{{km}^{-2}}$ and $B=83~{{MeV}/{{fm}^3}}$. Further, with the motivation to discuss the effects of the increasing value of the electric charge, we have predicted numerical values of the different physical parameters for the strange star candidate $LMC~X-4$ in Table~\ref{Table 2}. The high redshift value $(0.2824-0.2853)$ supports that the proposed model is suitable to study strange star candidates. Both Tables.~{\ref{Table 1} and \ref{Table 2} feature that due to different values of $\alpha$ the predicted values of mass to radius ratio for the different strange star candidates are well with in the upper limit of the mass-radius ratio provided by Andr{\'e}asson~\cite{Andreasson2009}. 

In a summery, in this article we have presented an anisotropic charged spherically symmetric stellar model which is suitable to study ultra-dense strange stars.

\section*{Acknowledgments}
SR and FR are thankful to the Inter-University Centre for Astronomy and Astrophysics (IUCAA), Pune, India for providing Visiting Associateship under which a part of this work was carried out. SR is also thankful to the authority of The Institute of Mathematical Sciences, Chennai, India for providing all types of working facility and hospitality under the Associateship scheme. FR is also thankful to DST-SERB (EMR/2016/000193), Govt. of India for providing financial support. A part of this work was completed while DD was visiting the IUCAA, Pune, India and the author gratefully acknowledges the warm hospitality and facilities at the library there. The work by MK was supported by Russian Science Foundation and carried in the framework of MEPhI Academic Excellence Project (contract 02.a03.21.0005, 27.08.2013).

\section*{Appendix}\label{App}
The expressions of the constants $\lambda_{{1}}$,~$\lambda_{{2}}$,~$\nu_{{1}}$,~$\nu_{{2}}$,~$\nu_{{3}}$,~$\nu_{{4}}$,~$\nu_{{5}}$~and~$\nu_{{6}}$ which are used in Eqs.~(\ref{4.1}) and ({\ref{4.2}}) given as
\begin{eqnarray}\label{App1}
&\qquad\hspace{-1cm} \lambda_{{1}}=-2\,{R}^{5}{\alpha}^{2}-16\,B\pi \,{R}^{3}+3\,M, \\ \label{App2}
&\qquad\hspace{-1cm} \lambda_{{2}}=3\,{R}^{7}{\alpha}^{2}+16\,B\pi \,{R}^{5}-5\,M{R}^{2},\\ \label{App3}
&\qquad\hspace{-1cm} \nu_{{1}}=\Big( 9\,{R}^{10}{\alpha}^{4}+96\,B\pi \,{R}^{8}{\alpha}^{2}+256\,
{B}^{2}{\pi }^{2}{R}^{6}\nonumber\\
&\qquad\hspace{0cm}-30\,M{R}^{5}{\alpha}^{2}+8\,{R}^{6}{\alpha}^{2}-160\,BM\pi \,{R}^{3}+64\,B\pi \,{R}^{4}\nonumber\\
&\qquad\hspace{2cm}+25\,{M}^{2}-12\,MR\Big)^{1/2},\\ \label{App4}
&\qquad\hspace{0cm} \nu_{{2}}=\Big[{\frac {9}{512}}{R}^{10}{\alpha}^{4}+\frac{1}{4}B\pi \,{R}^{8}{
\alpha}^{2}+{B}^{2}{\pi }^{2}{R}^{6}-{\frac {3}{64}}M{R}^{5}{\alpha}^{2}\nonumber\\
&\qquad\hspace{3cm}-{\frac {11}{32}}B M\pi {R}^{3}+{\frac {15}{512}}{M}^{2}\Big],\\ \label{App5}
&\qquad\hspace{-1cm} \nu_{{3}}=\frac{1}{16}{R}^{5}{\alpha}^{2}+\frac{2}{3}B\pi {R}^{3}-\frac{M}{8},\\ \label{App6}
&\qquad\hspace{-1cm} \nu_{{4}}=2\,{R}^{5}{\alpha}^{2}+16\,B\pi \,{R}^{3}-3 M,\\ \label{App7}
&\qquad\hspace{-1cm} \nu_{{5}}=-3\,{R}^{7}{\alpha}^{2}-16\,B\pi \,{R}^{5}+5\,M{R}^{2},\\ \label{App8}
&\qquad\hspace{-1cm} \nu_{{6}}=\Big[\nu_{{3}}\ln  \left(2 M -{R}^{5}{\alpha}^{2}-R \right) +\frac{1}{16}\nu_{{4}}\ln  \big( {Q}^{2}+{R}^{2}\nonumber\\
&\qquad\hspace{1cm}-2 MR \big) +\frac{2}{3} \left( B\pi \,{R}^{3}-\frac{3}{16} M \right) \ln  \left( R \right)\Big].
\end{eqnarray}


\begin{thebibliography}{999}
	
\bibitem{Itoh1970} N. Itoh, Prog. Theor. Phys. \textbf{44}, 291 (1970)

\bibitem{Farhi1984} E. Farhi, R.L. Jaffe, Phys. Rev. D \textbf{30}, 2379 (1984)

\bibitem{Alcock1986} C. Alcock, E. Farhi, A. Olinto, Astrophys. J. \textbf{310}, 261 (1986)

\bibitem{Haensel1986} P. Haensel, J.L. Zdunik, R. Schaefer, Astron. Astrophys. \textbf{160}, 121 (1986)

\bibitem{Bodmer1971} A.R. Bodmer, Phys. Rev. D \textbf{4}, 1601 (1971)

\bibitem{Witten1984} E. Witten, Phys. Rev. D \textbf{30}, 272 (1984)

\bibitem{Terazawa} H. Terazawa, INS, Univ. of Tokyo Report No. INS-Report-338, 1979; J. Phys. Soc. Jpn. \textbf{58}, 3555 (1989); \textbf{58}, 4388 (1989); \textbf{59}, 1199 (1990).

\bibitem{Alcock1988} C. Alcock, A.V. Olinto, Annu. Rev. Nucl. Part. Sci. \textbf{38}, 161 (1988)

\bibitem{Madsen1999} J. Madsen, Lect. Notes Phys. \textbf{516}, 162 (1999)

\bibitem{Glendenning1995a} N.K. Glendenning, Ch. Kettner, F. Weber, Astrophys. J. \textbf{450}, 253 (1995)

\bibitem{Glendenning1995b} N.K. Glendenning, Ch. Kettner, F. Weber, Phys. Rev. Lett. \textbf{74}, 3519 (1995)

\bibitem{Kettner1995} Ch. Kettner, F. Weber, M.K. Weigel, N.K. Glendenning, Phys. Rev. D \textbf{51}, 1440 (1995)

\bibitem{Weber2005} F. Weber, Prog. Part. Nucl. Phys. \textbf{54}, 193 (2005)

\bibitem{Perez-Garcia2010} M.A. Perez-Garcia, J. Silk, J.R. Stone, Phys. Rev. Lett. \textbf{105}, 141101 (2010)

\bibitem{Rodrigues2011} H. Rodrigues, S.B. Duarte, J.C.T. de Oliveira, Astrophys. J. \textbf{730}, 31 (2011)

\bibitem{Bordbar2011} G.H. Bordbar, A.R. Peivand, Res. Astron. Astrophys. \textbf{11}, 851 (2011)

\bibitem{Usov2004} V. Usov, Phys. Rev. D \textbf{70}, 067301 (2004)

\bibitem{Usov2005} V. Usov, T. Harko, K.S. Cheng, Astrophys. J. \textbf{620}, 915 (2005)

\bibitem{Ray2003} S. Ray, A.L. Esp{\'i}ndola, M. Malheiro, J.P.S. Lemos, V.T. Zanchin, Phys. Rev. D \textbf{68}, 084004 (2003)

\bibitem{Siffert2007} B.B. Siffert, J.R. de Mello, M.O. Calv{\~a}o, Braz. J. Phys. \textbf{37}, 2B (2007)

\bibitem{Negreiros2009} R.P. Negreiros, F. Weber, M. Malheiro, V. Usov, Phys. Rev. D \textbf{80}, 083006 (2009)

\bibitem{Varela2010} V. Varela, F. Rahaman, S. Ray, K. Chakraborty, M. Kalam, Phys. Rev. D \textbf{82}, 044052 (2010)

\bibitem{Malheiro2011} M. Malheiro, R.P. Negreiros, F. Weber, V. Usov, J. Phys. Conf. Ser. \textbf{312}, 042018 (2011)

\bibitem{Rahaman2012} F. Rahaman, R. Sharma, S. Ray, R. Maulick, I. Karar, Eur. Phys. J. C \textbf{72}, 2071 (2012)

\bibitem{Arbanil2013} J.D.V. Arba{\~a}il, J.P.S. Lemos, V.T. Zanchin, Phys. Rev. D \textbf{88}, 084023 (2013)

\bibitem{Sunzu2014} J.M. Sunzu, S.D. Maharaj, S. Ray, Astrophys. Space Sci. \textbf{352}, 719 (2014)

\bibitem{Arbanil2015} J.D.V. Arba{\~a}il, M. Malheiro, Phys. Rev. D \textbf{92}, 084009 (2015)

\bibitem{Panahi2016} H. Panahi, R. Monadi, I. Eghdami, Chin. Phys. Lett. \textbf{33}, 072601 (2016) 

\bibitem{Ruderman1972} R. Ruderman, Rev. Astron. Astrophys. \textbf{10}, 427 (1972) 

\bibitem{Bowers1974} R.L. Bowers, E.P.T. Liang, Class. Astrophys. J. \textbf{188}, 657 (1974) 
 
\bibitem{Herrera1997} L. Herrera, N.O. Santos, Phys. Report. \textbf{286}, 53 (1997)

\bibitem{Dev2002} K. Dev, M. Gleiser, Gen. Relativ. Gravit. \textbf{34}, 1793 (2002)

\bibitem{Dev2003} K. Dev, M. Gleiser, Gen. Relativ. Gravit. \textbf{35}, 1435 (2003)

\bibitem{Dev2004} M. Gleiser, K. Dev, Int. J. Mod. Phys. D \textbf{13}, 1389 (2004)

\bibitem{Ivanov2002} B.V. Ivanov, Phys. Rev. D \textbf{65}, 104011 (2002)

\bibitem{SM2003} F.E. Schunck, E.W. Mielke, Class. Quantum Gravit. \textbf{20}, 301 (2003)

\bibitem{MH2003} M.K. Mak, T. Harko, Proc. R. Soc. A \textbf{459}, 393 (2003)

\bibitem{Varela2010} V. Varela, F. Rahaman, S. Ray, K. Chakraborty, M. Kalam, Phys. Rev. D \textbf{82}, 044052 (2010)

\bibitem{Rahaman2010} F. Rahaman, S. Ray, A.K. Jafry, K. Chakraborty, Phys. Rev. D \textbf{82}, 104055 (2010)

\bibitem{Rahaman2011} F. Rahaman, P.K.F. Kuhfittig, M. Kalam, A.A. Usmani, S. Ray, Class. Quantum Gravit. \textbf{28}, 155021 (2011)

\bibitem{Rahaman2012} F. Rahaman, R. Maulick , A.K. Yadav, S. Ray, R. Sharma, Gen. Relativ. Gravit. \textbf{44}, 107 (2012)

\bibitem{Kalam2012} M. Kalam, F. Rahaman, S. Ray, Sk.M. Hossein, I. Karar, J. Naskar, Eur. Phys. J. C \textbf{72}, 2248 (2012)

\bibitem{Maurya2016} S.K. Maurya, Y.K. Gupta, S. Ray, D. Deb, Eur. Phys. J. C \textbf{76}, 693 (2016)

\bibitem{Maurya2017} S.K. Maurya, D. Deb, S. Ray, P.K.F. Kuhfittig, arXiv:1703.08436[physics.gen-ph]

\bibitem{Deb2017} D. Deb, S. R. Chowdhury, S. Ray, F. Rahaman, B.K. Guha, Ann. Phys. \textbf{387}, 239 (2017) 

\bibitem{Ovalle2017a} J. Ovalle, Phys. Rev. D {\bf 95}, 104019 (2017)

\bibitem{Ovalle2017b} J. Ovalle, R. Casadio, R. da Rocha, A. Sotomayor, arXiv:1708.00407v2 [gr-qc]  (To appear in EPJC)

\bibitem{Chodos1974} A. Chodos, R.L. Jaffe, K. Johnson, C.B. Thorn, V.F. Weisskopf, Phys. Rev. D \textbf{9}, 3471 (1974)

\bibitem{Brilenkov2013} M. Brilenkov, M. Eingorn, L. Jenkovszky, A. Zhuk, JCAP \textbf{08}, 002 (2013)

\bibitem{Panda2015} N.R. Panda, K.K. Mohanta, P.K. Sahu, J. Physics: Conference Series \textbf{599}, 012036 (2015)

\bibitem{Isayev2015} A.A. Isayev, Phys. Rev. C \textbf{91}, 015208 (2015)

\bibitem{Maharaj2014} S.D. Maharaj, J.M. Sunzu, S. Ray, Eur. Phys. J. Plus. \textbf{129}, 3 (2014)

\bibitem{Paulucci2014} L. Paulucci, J.E. Horvath, Phys. Lett. B \textbf{733}, 164 (2014)

\bibitem{Abbas2015}  G. Abbas, S. Qaisar, A. Jawad, Astrophys. Space Sci. \textbf{359}, 57 (2015)

\bibitem{Arbanil2016} J.D.V. Arba{\~n}il, M. Malheiro, JCAP \textbf{11}, 012 (2016)

\bibitem{Lugones2017} G. Lugones, J.D.V. Arba{\~n}il, Phys. Rev. D \textbf{95}, 064022 (2017)

\bibitem{Rahaman2014} F. Rahaman, K. Chakraborty, P. K. F. Kuhfittig, G. C. Shit and M. Rahman, Eur. Phys. J. C  \textbf{74}, 3126 (2014)

\bibitem{Felice1995} F. de Felice, Y.Q. Yu, J. Fang, Mon. Not. R. Astron. Soc. \textbf{277}, L17 (1995)

\bibitem{Felice1999} F. De Felice, S.M. Liu, and Y.Q. Yu, Class. Quantum Gravit. \textbf{16}, 2669 (1999)

\bibitem{Tolman1939} R.C. Tolman, Phys. Rev. \textbf{55}, 364 (1939)

\bibitem{Oppenheimer1939} J.R. Oppenheimer, G.M. Volkoff, Phys. Rev. \textbf{55}, 374 (1939)

\bibitem{Dionysiou1982} D.D. Dionysiou, Astrophys. Space Sci. {\bf 85}, 331 (1982)

\bibitem{Harko2002} M.K. Mak, T. Harko,  Chin. J. Astron. Astrophys. \textbf{2}, 248 (2002)

\bibitem{Buchdahl1959} H.A. Buchdahl, Phys. Rev. D \textbf{116}, 1027 (1959)

\bibitem{Andreasson2009} H. Andr{\'e}asson, Commun. Math. Phys. \textbf{288}, 715 (2009)

\bibitem{Herrera1992} L. Herrera, Phys. Lett. A \textbf{165}, 206 (1992)

\bibitem{Abreu2007} H. Abreu, H. Her{\'n}andez, L.A. N{{\'u}}${\tilde{n}}$ez, Class. Quantum Grav. \textbf{24}, 4631 (2007)

\bibitem{demorest} P.B. Demorest, T. Pennucci, S.M. Ransom, M.S.E. Roberts, J.W.T. Hessels, Nature {\bf 467}, 1081 (2010)

\bibitem{dey2013} T. Gangopadhyay, S. Ray, X.-D. Li, J. Dey, M. Dey, Mon. Not. R. Astron. Soc. {\bf 431}, 3216 (2013)

\bibitem{guver2010b} T. G{\"u}ver, P. Wroblewski, L. Camarota, F. {\"O}zel, Astrophys J. {\bf 712}, 964 (2010)

\bibitem{guver2010a} T. G{\"u}ver, P. Wroblewski, L. Camarota, F. {\"O}zel, Astrophys J. {\bf 719}, 1807 (2010)

\bibitem{ozel2009} F. {\"O}zel, T. G{\"u}ver, D. Psaltis, Astrophys J. {\bf 693}, 1775 (2009)

\end{thebibliography}
\end{document}